\documentclass[12pt]{article}
\usepackage[svgnames]{xcolor}
\usepackage{amsmath,amsfonts,amssymb}
\usepackage{cite}
\usepackage{graphicx}
\usepackage{hyperref}
\usepackage{multirow}
\hypersetup{
linktocpage=true,
colorlinks=true,
linkcolor=blue,
citecolor=blue,
urlcolor=magenta,
pdfpagemode=UseOutlines
}
\usepackage{fullpage}
\usepackage{array,longtable}
\usepackage[table]{xcolor}
\usepackage{colortbl}
\usepackage{comment}
\renewcommand{\arraystretch}{1.8} 
\usepackage{eso-pic}

\makeatletter
\newcommand{\@preprint}{}
\newcommand{\preprint}[1]{\gdef\@preprint{#1}}
\newlength{\preprinttop}
\newlength{\preprintright}
\let\articlemaketitle\maketitle
\renewcommand{\maketitle}{%
  \ifx\@preprint\@empty\else
    \AddToShipoutPictureFG*{%
      \AtPageUpperLeft{%
        \put(
          \LenToUnit{\dimexpr\paperwidth-\preprintright\relax},
          \LenToUnit{\dimexpr-\preprinttop\relax}
        ){%
        \makebox[0pt][r]{%
  \normalfont\small
  \shortstack[r]{\@preprint}%
}%
        }%
      }%
    }%
  \fi
  \articlemaketitle
}
\makeatother

\graphicspath{{./figures/}}

\let\Re\undefined
\let\Im\undefined
\DeclareMathOperator{\Re}{Re}
\DeclareMathOperator{\Im}{Im}

\usepackage{authblk}

\preprint{LA-UR-26-29844\\MIT-CTP/6147}
\title{\vspace{-2em}Machine learning for lattice QCD}

\author[1]{Gurtej Kanwar}
\author[2]{Scott Lawrence}
\author[3]{Alessandro Nada}
\author[4,5]{Phiala Shanahan}

\affil[1]{Higgs Centre for Theoretical Physics, The University of Edinburgh, James Clerk Maxwell Building, Peter Guthrie Tait Road, Edinburgh, EH9 3FD, UK}

\affil[2]{Theoretical Division,
Los Alamos National Laboratory,
Los Alamos, NM 87545, USA}

\affil[3]{Dipartimento di Fisica,
Universit\`a degli Studi di Torino and INFN, Sezione di Torino, Via Pietro Giuria 1, 10125 Turin, Italy}

\affil[4]{Center for Theoretical Physics -- a Leinweber Institute,
Massachusetts Institute of Technology,
Cambridge, MA 02139, USA}
\affil[5]{The NSF AI Institute for Artificial Intelligence and Fundamental Interactions}

\usepackage{tocloft}
\begin{document}

\maketitle
\begin{abstract}
Machine learning (ML) methods have been explored as a means of optimizing lattice field theory calculations, with increasing interest in recent years. This chapter gives an introduction and summary to the use of ML methods to optimize lattice field theory without introducing any uncontrolled approximations. We focus on those directions which are clearly established as of late 2025, marking the 50th anniversary of the beginning of the study of lattice gauge theory.
This chapter is a contribution to the online book ``Lattice QCD at 50 years,'' (LQCD@50), edited by Tanmoy Bhattacharya, Maarten Golterman, Rajan Gupta, Laurent Lellouch, and Steve Sharpe. 
\end{abstract}

\tableofcontents


\section{Introduction}

Machine learning (ML) is a branch of artificial intelligence that enables pattern identification and the development of data-driven predictions or decision-making. Many applications of ML are based on neural networks, which are highly flexible models, typically with huge numbers of free parameters, that are capable of fitting high-dimensional, nonlinear functions and are particularly well-suited for tasks such as interpolation, regression, and the emulation of computationally expensive operations. In scientific computing, ML techniques can be leveraged to optimize algorithms, accelerate simulations, and uncover structures in complex data; see Refs.~\cite{Mehta:2018dln, raschka2022machine,Marquardt2025MLPhysicists} for accessible introductions to ML.

In Lattice Quantum Field Theory (LQFT), ML has now been applied (with varying levels of maturity and/or success) to accelerate every part of the calculational workflow, including: defining lattice actions, sampling gauge field configurations, accelerating solvers, observable estimation, and aspects of analysis~\cite{Boyda:2022nmh}. 
A separation can be made between applications of ML in LQFT that are {\it model-based} (e.g., efforts to define order parameters, the use of neural networks in fits and parameterisations of data, and applications to inverse problems), and those that {\it demand exactness guarantees} in the sense of asymptotic correctness. This is an important distinction; unlike many areas of applied ML where approximations and heuristic or `black box' methods are acceptable, LQFT is inherently a first-principles approach: therefore, integration of ML techniques into many aspects of the LQFT workflow must be undertaken in such a way that any mathematical guarantees afforded by conventional algorithms are preserved. 
This category of applications---those where unbiasedness must be guaranteed---is the focus of this chapter.

Methods of accelerating computations through ML while maintaining the mathematical rigour of traditional approaches can be broadly grouped into three classes based on the role of ML within the computational workflow:
\begin{itemize}
    \item {\it Optimise existing algorithms}: One straightforward strategy is to accelerate existing computations by tuning algorithmic parameters using ML. This approach preserves the structure and logic of the original algorithm while enhancing its performance. For instance, free parameters in algebraic multigrid methods for solving linear systems (e.g., for propagator inversions) can be optimized using ML techniques. Similarly, replacing stochastic optimization approaches in various contexts with automatic differentiation can lead to performance improvements without sacrificing exactness. 
    \item {\it Transform the problem:} A second strategy involves using ML to transform the problem into an equivalent formulation that is computationally more tractable to solve than the original. Standard algorithmic approaches, with the associated guarantees, are then applied to the transformed problem. This includes techniques such as preconditioning, which can significantly improve the convergence rate of numerical solvers, and change of variables—for example, deforming the integration contour—which can reduce the variance of numerical integration via Monte Carlo without altering the central value. 
    \item {\it Solve a related problem and correct:} A third approach leverages ML to solve a related but more accessible problem and then apply a known correction to achieve the exact solution. One example includes learning a mapping between a low-precision and a high-precision solution in linear systems and subsequently correcting for the introduced bias. Another example involves sampling from a nearby probability distribution followed by a reweighting or resampling step to obtain the correct distribution. This method allows for flexibility in computational methodology while ensuring that the final results remain provably exact.
\end{itemize}
Table~\ref{tab:classification} collects a (non-exhaustive) grid of these three classes of applications of asymptotically-correct ML-accelerated approaches to the different steps of the LQFT workflow. Some applications are in the early stages of investigation at the current time, while others are approaching maturity. None are, as of late 2025, yet the default approach in state-of-the-art LQFT calculations at scale.

An important reason that ML approaches are still far from being default in LQFT applications is the wide gap between the standard ML/AI workflow and the specialized workflows and infrastructure that have developed within LQFT. Most ML research and engineering is carried out in Python, using a small set of powerful toolkits—most prominently TensorFlow~\cite{tensorflow2015-whitepaper}, PyTorch~\cite{paszke2019pytorchimperativestylehighperformance}, and JAX~\cite{jax2018github}. These libraries provide highly optimized routines for defining neural network architectures and performing gradient-based optimization (e.g., through backpropagation). They have been designed to lower the barrier to entry, allowing rapid prototyping and flexible experimentation for applications close to typical ML use cases. 
By contrast, the computational infrastructure in LQFT has been shaped by decades of large-scale scientific computing. LQFT calculations are typically performed using custom-built software packages such as Chroma~\cite{Edwards:2004sx}, CPS~\cite{Jung:2014ata}, or QUDA~\cite{Clark:2009wm}, which have been engineered to exploit the architectures of high-performance computing (HPC) platforms. These codes are tuned for massively parallel execution on distributed-memory supercomputers, optimized for efficient communication over interconnects, and adapted to exploit GPU acceleration at scale. Every layer of the software stack has been refined to address the specific numerical challenges of simulating gauge field configurations, propagators, and correlators in gauge theories.

Bringing ML methods into this environment introduces substantial technical challenges. Standard neural network architectures—such as convolutional or transformer-based models—are not naturally aligned with the data structures or symmetry properties of lattice gauge theory. In typical applications of ML, the conventional wisdom (the ``bitter lesson''~\cite{sutton2019bitter}) is that incorporating symmetries is often not necessary or even helpful. The (gauge) symmetries of lattice QCD, however, are extremely high-dimensional, and anecdotal evidence to date strongly indicates that it is worth including manifest gauge (in/co)variance in the design of networks being trained~\cite{Kanwar:2020xzo}.
Symmetry-respecting neural networks have been developed to address this challenge~\cite{Favoni:2020reg}: by explicitly encoding gauge invariance or rotational symmetries into the architecture, these models can capture the essential structure of the problem in a way that black-box networks cannot. These applications, however, are specialized, and scaling custom models efficiently across tens of thousands of nodes or GPUs, with careful management of memory hierarchies and communication bottlenecks, requires significant re-engineering. That is, the high-throughput, distributed training strategies common in commercial ML do not map easily onto the specialized HPC workflows that define state-of-the-art lattice calculations. Thus, even when an ML model shows promise in small-scale exploratory studies, extending it to production-level simulations demands overcoming both algorithmic and systems-level hurdles.

The long-term challenge, then, is twofold: first, to continue developing specialized ML architectures that respect the mathematical structure of gauge theories, and second, to integrate them into HPC environments at a level of scalability and efficiency comparable to traditional lattice methods. Meeting both demands is essential if ML is to become a true complement—or even a replacement—in the established LQFT computational toolkit. Despite the magnitude of these challenges, the rapid pace of innovation at the intersection of ML and physics gives strong reason for optimism that ML will ultimately become an integral part of the future of LQFT.



\newcommand{\sqcell}[1]{\cellcolor{white}\parbox[c][3.2cm][c]{3.2cm}{\centering #1}}

\begin{table}
\begin{center}

{
\renewcommand{\arraystretch}{1.2}
\begin{tabular}{c l l}
\hline
\hline \\[-.5ex]
     \bf Approach & \multicolumn{1}{c}{\bf ML application} & \multicolumn{1}{c}{\bf Area}  \\[2ex]
\hline \\[-.5ex]
    \multirow{4}{4.0cm}{\centering Optimize existing algorithm}
    & Correlated sampling & Obs.\ estimation \\
    & Tune MG params. & \begin{tabular}{@{}l@{}}Config.\ generation, \\[-1ex] Obs.\ estimation\end{tabular} \\
    & Learned interpolators & Obs.\ estimation \\
    & Perfect actions & Action \\[2ex]
\hline \\[-.5ex]
    \multirow{3}{4.0cm}{\centering Transform the problem}
    & Learned-manifold HMC & Config.\ generation \\
    & Learned preconditioners & Solvers \& inversion \\
    & Contour deformation & \begin{tabular}{@{}l@{}}Obs.\ estimation, \\[-1ex] Action \end{tabular} \\[3ex]
\hline \\[-.5ex]
    \multirow{2}{4.0cm}{\centering Solve related problem (and correct)}
    & Generative models &  Config.\ generation \\
    & Surrogate observables & Obs.\ estimation \\[2ex] 
\hline
\hline
\end{tabular}
}
\end{center}
\caption{\label{tab:classification} Classification of some ML applications to LQFT based on the part of the computational pipeline that is targeted, and the approach to maintaining exactness. Table is not exhaustive.}

\end{table}

The remainder of this chapter is divided into four sections, with a brief introduction to ML followed by the ML approaches to LQFT categorized depending on the manner in which the lattice calculation is accelerated or improved. If the main computational obstacle is performing importance sampling---that is, if the observable of interest is fast to evaluate on any given sample---then using machine learning methods to accelerate the Monte Carlo sampling itself is of greatest value. This class of approaches, exemplified by normalizing flows, is discussed in Section~\ref{sec:mc-acceleration}. If the total number of samples on which an observable is to be estimated is held fixed, then ML-based acceleration must improve the signal-to-noise ratio by modifying the observable itself (e.g.~via control variates); this is discussed in Section~\ref{sec:estimators}. Finally, if there are a large number of samples available but evaluating the observable is itself computationally challenging, then ML may be used to accelerate (or approximate) the evaluation of the observable; this is discussed in Section~\ref{sec:acceleration}.

\section{Brief introduction to ML/AI}

Here we include a brief introduction to machine learning methods, with a bias towards the points of view that have been most productive in computational physics. This introduction is not intended to be comprehensive for a reader entirely new to the field. We recommend such readers refer to the introductory material for PyTorch, which provides a good practical introduction to ML, or Refs.~\cite{Mehta:2018dln, raschka2022machine,Marquardt2025MLPhysicists}.

In great generality, many calculations may be accelerated or improved if we have numerical access to some function, which has some desired mathematical property, and which can be evaluated sufficiently quickly. ``Machine learning'' is the name given to the process of searching for such a function.

To have a concrete example, we might try to learn a function $f(U)$ which does a good job of approximating the action $S(U)$. This would be useful when the action is slow to evaluate, for example, and is not too different from some examples we will see later in this text. We want our function $f(U)$ to be as close as possible to $S(U)$, and we want it to be fast to evaluate.

The paradigm of machine learning proceeds in the following steps:
\begin{enumerate}
    \item Identify a family of functions, $f_\theta(U)$,
    defined by a set of real parameters $\theta$.
    \item Define a measure of `loss'; a functional $L[f_\theta]$ which is small when $f_\theta(\cdot)$ has our desired property.
    \item Perform some search algorithm---typically an elaboration on gradient descent---to find parameters $\theta$ such that $L(\theta) \equiv L[f_\theta]$ is as small as possible. This optimization step is often referred to as ``training''.
    \item Employ the resulting $f_\theta(\cdot)$ in a calculation.
\end{enumerate}
In the first step we must identify a family of functions. The reader will not be surprised that this family is often defined by a neural network: the weights constitute the parameters $\theta$, and the gauge configuration $U$ shall be passed as an input. The most straightforward neural network is the \emph{multi-layer perceptron}: an alternating composition of element-wise nonlinear transformations with arbitrary linear transformations\footnote{Sparser architectures are possible and in fact common, but the fundamental idea---that element-wise nonlinear transformations and fairly arbitrary linear transformations are both fast---recurs.}. In the case of one hidden layer (meaning one nonlinear transformation, sandwiched by arbitrary matrix multiplications), this may be written:
\begin{equation}
    \mathrm{NN}(x)_i =
    M^{(2)}_{ij} \sigma\left(M^{(1)}_{jk} x_k\right)
    \text,
\end{equation}
where the function $\sigma : \mathbb R \rightarrow \mathbb R$ is 
a nonlinear function, termed an ``activation function''.\footnote{Typical examples are $\sigma(x) = \tanh(x)$ or $\sigma(x) = \max(0, x)$.}

Why a neural network? Historically, neural networks were justified by somewhat stretched analogies to the structure of the nervous systems of living things. Today it is probably more accurate to simply observe that, because they rely so heavily on linear algebra, neural networks are an effective mechanism for packing a large number of free parameters into a relatively small amount of computation. Moreover, the computations that are involved map nicely onto modern computing architectures, such as GPU-based machines.

In the second step we must define a loss functional $L[f_\theta]$, which is also implicitly a loss \emph{function} $L(\theta)$ of the parameters. This loss function will typically be evaluated stochastically. In the case where we are seeking some function which approximates the action, a suitable loss functional might be
\[
L[f_\theta] = \int dU\, \left[S(U) - f_\theta(U)\right]^2\text,
\]
which may be approximated by Monte Carlo sampling from the measure $dU$.

It is in the third step that the computational part of the learning actually happens: we search for parameters $\theta$ that achieve a small loss $L(\theta)$. If $L(\cdot)$ has been chosen wisely, then achieving a small-but-not-minimal loss will still be helpful. Because $L(\cdot)$ is typically expressed as an integral and can only be evaluated by Monte Carlo, we perform \emph{stochastic} gradient descent or one of its relatives. All\footnote{Of course there are approximate minimization algorithms which depart from this paradigm. The most extreme examples are \emph{gradient-free} methods. These lie outside the scope of this chapter.} follow the same basic pattern:
\begin{equation}
    \theta_{n+1} = \theta_{n} - \epsilon \left(\frac{\partial L(\theta)}{\partial \theta}\right)\text.
\end{equation}
In stochastic gradient descent, $\epsilon$ is a fixed number (termed the ``learning rate'') and the partial derivative is evaluated by Monte Carlo as discussed above. In more elaborate methods, including the widely used \texttt{Adam}~\cite{kingma2014adam}, the estimate of the gradient and/or the learning rate is modified, taking into account the history of the gradients (e.g.~to add momentum, or estimate local curvature).

Much of the philosophy behind modern machine learning and AI is reflected in the following heuristic: having more parameters, and spending more time training, is what ultimately drives a successful minimization of the loss. These factors have historically overwhelmed advanced ans\"atze or clever minimization methods. This observation is sometimes called the ``bitter lesson''~\cite{sutton2019bitter}. This does not imply that no thought is required in machine learning: there is no escaping the need to select a meaningful loss function, for example. But it does provide a guide. The purpose of the methods in this chapter is, by and large, to find new ways of bringing a great deal of computational power to bear on specific aspects of LQCD.

\section{Monte Carlo acceleration}\label{sec:mc-acceleration}

The fundamental problem in lattice QCD is calculating thermal expectation values via the Euclidean path integral
\begin{equation}
\label{eq:path_integral}
  \langle \mathcal{O} \rangle = \frac{1}{Z} \int \mathcal{D} U \, e^{-S(U)}  \mathcal{O}(U)
\end{equation}
with $\mathcal{O}$ an operator to be measured and $Z$ being the partition function of the theory,
\begin{equation}
\label{eq:partition_function}
  Z = \int \mathcal{D} U \, e^{-S(U)}.
\end{equation}
This problem can be numerically solved by \emph{importance sampling} according to the Boltzmann distribution $e^{-S(U)}/Z$ defined by Eq.~\eqref{eq:path_integral}.
Given an ensemble of field configurations $\{ U^{(1)}, \dots, U^{(N)} \}$ sampled in this way, the sample mean provides a Monte Carlo estimate of the expectation value,
\begin{equation}
\label{eq:mc_obs_estimator}
    \langle \mathcal{O} \rangle \simeq \frac{1}{N} \sum_{i=1}^N \mathcal{O}(U^{(i)}).
\end{equation}
Increasing the number of samples $N$ systematically lowers the uncertainties on this estimate.

\subsection{Markov Chain Monte Carlo}
Markov Chain Monte Carlo algorithms provide a practical way to perform the importance sampling above by applying a memoryless sequence of stochastic updates
\begin{equation}
    U^{(1)} \to U^{(2)} \to U^{(3)} \to \dots
\end{equation}
that are constructed to asymptotically converge to the Boltzmann distribution.

\begin{figure}[t]
    \centering
    \includegraphics{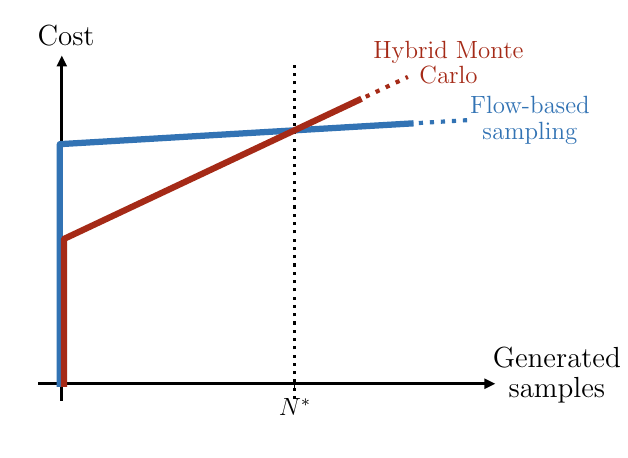}
    \caption{Schematic comparison of the sampling cost using a traditional Monte Carlo sampler, such as Hybrid Monte Carlo, versus an optimized flow-based sampler. The up-front cost of training a flow-based sampler is typically larger than the cost of thermalizing a Markov Chain sampler, while the cost per uncorrelated sample is typically smaller for a flow-based sampler. If the number of required samples exceeds the threshold $N^*$, the overall cost of sampling is lower when using the flow-based sampling strategy.}
    \label{fig:cost-comparison}
\end{figure}

The overall cost of producing $N$ independent, thermalized field configurations over which the average of Eq.~\eqref{eq:mc_obs_estimator} is calculated can be expressed as the cost $C_{\rm th}$ of the burn-in (or thermalization) of the Markov Chain plus the cost $N C_{\rm s}$ of sampling the $N$ decorrelated configurations, such that the total cost is
\begin{equation}
\label{eq:mc_cost_total}
 C_{\rm tot} (N) = C_{\rm th} + N C_{\rm s},
\end{equation}
as shown in Figure~\ref{fig:cost-comparison}. The cost per sample $C_{\rm s}$ is determined by a combination of the computational cost of individual Markov chain updates and the number of updates required to obtain an independent sample.

Hybrid Monte Carlo~\cite{Duane:1987de} updates are the workhorse of lattice QCD simulations, based on integrating fictitious Hamiltonian dynamics to propose the next configuration in the Markov chain,
\begin{equation}
    U^{(i-1)} \xrightarrow[\text{Ham.\ dyn.}]{} U'.
\end{equation}
In Hybrid Monte Carlo, as with many Markov chain methods, this proposal must be corrected with a Metropolis-Hastings step which accepts the proposal for the next state in the chain with probability
\begin{equation}
    A(U^{(i-1)}, U') = \min \left( 1 , e^{-\Delta H} \right).
\end{equation}
Here the integrator error determines the violation $\Delta H$ of energy conservation in the integrated dynamics and therefore the acceptance rate and efficiency of the calculation.

\subsubsection{Critical slowing down}


For a given observable $\mathcal{O}$ and at fixed cost per Markov chain update, the cost $C_s$ per independent (thermalized) sample of $\mathcal{O}(U)$ typically scales as
\begin{equation}
\label{eq:mc_cost_sampling}
C_{\rm s} \propto 2\tau_{\rm int} (\mathcal{O}) \propto a^{-z}.
\end{equation}
The scaling of this cost is set by the integrated autocorrelation time $\tau_{\rm int} (\mathcal{O})$,
which measures the correlations in the measured observables $\mathcal{O}$ between sequential field configurations in the Markov chain~\cite{Wolff:2003sm}. 
It is well known that $\tau_{\rm int} (\mathcal{O})$ diverges as the lattice spacing $a$ approaches zero, or equivalently as the parameters are tuned towards a critical point where the correlation length diverges. In particular, when taking the continuum limit it diverges at least polynomially (as in Eq.~\eqref{eq:mc_cost_sampling}), with an exponent $z$ depending on the observable $\mathcal{O}$ and on the specific MCMC algorithm used during the simulation~\cite{Wolff:1989wq}. When sampling topological observables, the slowing down is particularly severe, with the exponent $z$ being particularly large, in a phenomenon denoted as topological freezing~\cite{DelDebbio:2004xh, Schaefer:2010hu, Hasenbusch:2017unr, Bonanno:2020hht, Eichhorn:2023uge}. Let us stress that this is an issue of Monte Carlo algorithms based on local updates and not of the observables themselves. 

\subsection{New sampling algorithms}
\label{sec:alternative_sampling}

Sampling algorithms that avoid or modify traditional MCMC approaches have been developed with the explicit goal of mitigating the rapid increase in costs indicated in Eq.~\eqref{eq:mc_cost_sampling}.
We expect these methods to be computationally cheaper than traditional MCMC methods only in a certain regime determined by the total start-up time (e.g., training time for ML methods) and the cost $C_s$ per independent sample. For complex ML architectures, for example, the training costs may easily exceed the thermalization costs, providing an advantage only when sufficiently many configurations need to be drawn. A qualitative picture of this situation is shown in Figure~\ref{fig:cost-comparison}, and identifying the associated cross-over in advantage may be useful as methods move towards practical applications.
In this section we provide a general description of some of these approaches, in particular those in which ML tools have the potential to be highly disruptive. We will describe the implementation of ML tools for flow-based approaches in section~\ref{sec:gen_models}.

\subsubsection{Learned MCMC dynamics}
One approach to circumventing the challenges of critical slowing down is to explicitly modify the MCMC dynamics to avoid slow-mixing modes as criticality is approached. For example, Fourier acceleration~\cite{Batrouni:1985jn} was proposed to mitigate critical slowing down for theories behaving similarly to a free scalar theory; however, obstacles remain in implementing this approach for gauge theories~\cite{Davies:1987vs,Duane:1988vr}. A generalization of this approach is to learn more arbitrary reparameterizations: implementations for Hybrid Monte Carlo dynamics have been investigated~\cite{Levy:2017bji,Foreman:2021ixr,Foreman:2021rhs,Foreman:2023ymy}.

\subsubsection{Learned actions}

Critical slowing down in MCMC simulations is generally related to the use of local updates. In fact, in the case of statistical models for which a global update is known, critical slowing down in the proximity of phase transitions is successfully mitigated~\cite{Wolff:1988uh,Swendsen:1987ce}. In general, though, the development of a global MCMC update is a highly non-trivial problem.

One way to design such an update is to introduce an effective action (or Hamiltonian) which can be efficiently simulated using global updates. In the Self-Learning Monte Carlo method~\cite{Huang:2017,Liu2017SLMC,Liu2016SLMCFermion,Xu:2017vug,Nagai2017SLMCCT,Shen2018SLMCDNN,Chen:2018aib,Liu:2019xnb,Nagai:2020ugq,Medvidovic:2020vum} the couplings of this effective action are learned through a training process using configurations generated with the original action. Then, a new MCMC is run by proposing new configurations using the effective action and accepting/rejecting them using the original one. In this way, a properly-trained effective long-range action can generate uncorrelated samples, without compromising exactness. A machine learning approach can be naturally applied in this setup employing an effective action parametrized by neural networks. In the case of lattice gauge theory, this has led to the construction of gauge-covariant neural networks~\cite{Nagai:2021bhh}; for further applications see refs.~\cite{Nagai:2023fxt,Nagai:2025rok}. 

Another intriguing approach to circumvent the issue of critical slowing down in lattice gauge theory is to use a new action characterized by strongly-suppressed lattice artifacts: this is the goal of the so-called fixed point (or classically-perfect) actions~\cite{Hasenfratz:1993sp,Blatter:1996np}, which are based on renormalization group transformations, and would allow better control of continuum extrapolations from simulations at coarser lattice spacings. A typical challenge when building this class of actions is finding a compact and accurate parametrization. This problem can be tackled in a new way using neural network-based architectures, which are much more expressive than the original approaches. In particular, recent efforts~\cite{Holland:2024muu,Holland:2025fsa} show how lattice gauge-equivariant convolutional neural networks (L-CNN)~\cite{Favoni:2020reg,Aronsson:2023rli} hold great promise in designing fixed-point actions in an efficient way.

\subsubsection{Trivializing flows}

An alternate approach is to transform the probability density to be sampled.
Taking the path integral of Eq.~\eqref{eq:path_integral}, if we perform a change of variables $V = f^{-1}(U)$, we obtain
\begin{equation}
    \langle \mathcal{O} \rangle = \frac{1}{Z} \int \mathcal{D} V \; \mathcal{O}(f(V)) \; e^{-S(f(V)) + \log |\det J_{f}(V)|}
\end{equation}
with $\det J$ being the determinant of the Jacobian of the transformation induced by $f$.
For this transformation to be well-defined, we assume $f$ is invertible and differentiable.
Now, the path integral can be computed by sampling from the new effective action $S_{\rm eff} (V) = S(f(V)) - \log |\det J_{f}(V)|$. If $S_{\rm eff}$ is constant, i.e., it corresponds to a uniform distribution, then $f$ is called a trivializing map~\cite{Luscher:2009eq}. 

\begin{figure}
    \centering
    \includegraphics{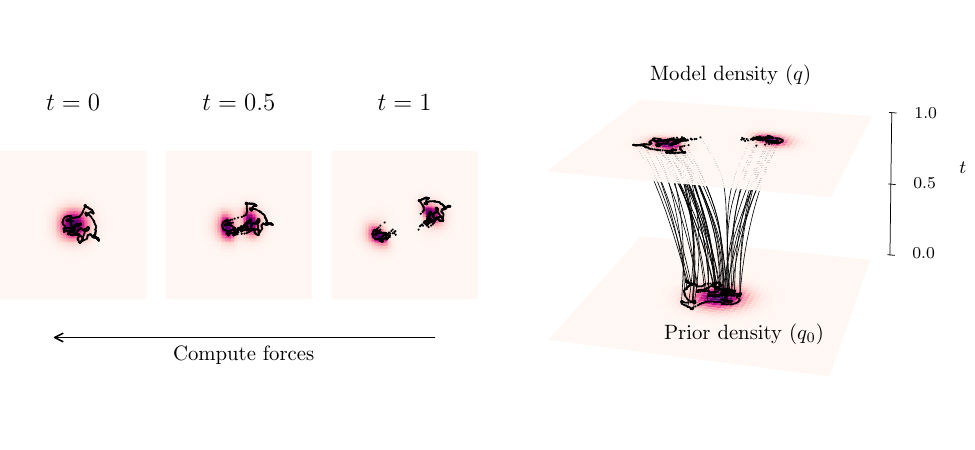}
    
    \includegraphics{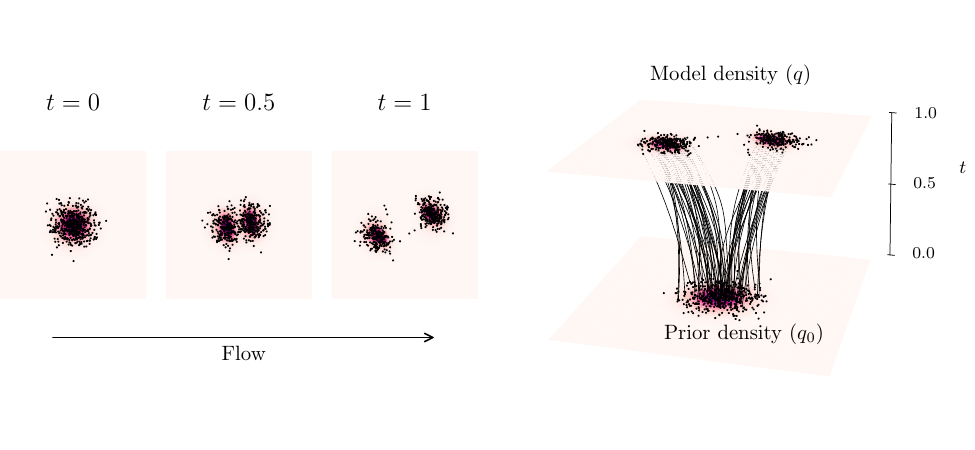}
    
    \caption{Comparison of applying a trivializing flow to improve HMC mixing time (top panel) versus using the same trivializing flow to perform direct importance sampling (bottom panel). The target model density is selected as a simple toy model with two widely separated modes, for which traditional sampling methods suffer from freezing.}
    \label{fig:flow_cartoons}
\end{figure}

Naturally, the challenge in this approach is to find a general non-perturbative method to find a true trivializing map $f$.
However, we stress that even an approximate trivializing map can effectively mitigate critical slowing down~\cite{Engel:2011re,Li:2018ihr}. The strategy of the simulation would follow the sketch in Figure~\ref{fig:flow_cartoons}: run a Monte Carlo using the new effective action and then flow back to the target distribution using $f$. If the approximate trivializing procedure is successful, smaller integrated autocorrelation times may be observed when running an MCMC. More precisely, this alternative sampling procedure may exhibit an improved scaling of the integrated autocorrelation time in the continuum, see Eq.~\eqref{eq:mc_cost_sampling}; however, we also stress that the scaling of the computational cost of an MCMC update using the new action has to be monitored as well. In section~\ref{sec:gen_models} we review some machine learning architectures that can be used to build the map $f$.

\subsubsection{Flow-based importance sampling}

A different but similar approach can be followed if the map $f$ is applied to field configurations $V$ sampled from a known, tractable distribution $q_0$. In this case, it is possible to explicitly write the probability distribution that the transformed configurations $U = f(V)$ follow as
\begin{equation}
\label{eq:is_probability}
    q (U) = q_0 ( f^{-1}(U) ) \left| \det J \right|^{-1},
\end{equation}
and expectation values $\left< \cdot \right>_q$ can easily be obtained by building an ensemble of samples $U$ as above.
Going back to the path integral of Eq.~\eqref{eq:path_integral}, one can then insert it
\begin{equation}
\label{eq:is_path_integral}
    \langle \mathcal{O} \rangle = \frac{\int \mathcal{D} U \, q(U) \, \left[ \frac{e^{-S(U)}}{q(U)}  \mathcal{O}(U) \right]}{\int \mathcal{D} U \, q(U) \, \left[ \frac{e^{-S(U)}}{q(U)} \right]}
\end{equation}
and use $q(U)$ for sampling. The factor
\begin{equation}
    w(U) = \frac{e^{-S(U)}}{q(U)}
\end{equation}
plays a crucial role, as its expectation value calculated with configurations sampled from $q(U)$ is the partition function of Eq.~\eqref{eq:partition_function}. A sketch of the structure of this sampling approach is shown in Figure~\ref{fig:flow_cartoons}.

Provided that $q$ and the target distribution $p$ are similar enough, there are essentially two alternatives in order to compute exactly the expectation values in Eq.~\eqref{eq:path_integral}.
\begin{itemize}
    \item The first possibility is to insert $w$ into the observable and compute the path integral with a reweighting procedure simply as
    \begin{equation}
    \label{eq:reweighting}
        \langle \mathcal{O} \rangle = \frac{\langle \mathcal{O}(U)  w(U) \rangle_{q}}{\langle w(U) \rangle_{q}}.
    \end{equation}
    The variance of this estimator clearly depends on the degree of overlap between $q(U)$ and $p(U)$. This can be monitored by looking at the Effective Sample Size (ESS), which is defined as
    \begin{equation}
        \text{ESS} = \frac{\langle w(U) \rangle_{q}^2}{\langle w(U)^2 \rangle_{q}}
    \end{equation}
    and goes from $0$ (vanishing overlap) to $1$ (perfect overlap).
    It can be seen as the asymptotic ratio $N / M$ relating the number of raw samples $M$ from $q(U)$ that must be drawn to achieve precision equivalent to $N$ independent samples from $p(U)$. Consequently, the average sampling cost of each independent  sample in this case can then be (approximately) expressed as
    \begin{equation}
    \label{eq:is_cost_sampling}
        C_{\rm s}^{(\rm flow)} \propto \frac{1}{\text{ESS}}.
    \end{equation}
    \item The second possibility is to build a new Markov Chain using samples from $q$ as proposals, 
    with a Metropolis-Hastings acceptance probability
    \begin{equation} \label{eq:flow-mcmc}
        A(U^{(i-1)}, U') = \min \left( 1 , \frac{w(U')}{w(U^{(i-1)})} \right) 
    \end{equation}
    to guarantee asymptotic convergence to the distribution $p(U)$.
    The sampling costs in this case still follow Eq.~\eqref{eq:mc_cost_sampling}, but the integrated autocorrelation time now becomes purely a measure of the overlap of $q(U)$ with the target. 
\end{itemize}

Generally speaking, finding a suitable transformation $f$ such that the overlap between $q(U)$ and the target $p(U)$ is sufficient represents a very difficult task. However, machine learning methods have begun to make promising inroads, as outlined in section~\ref{sec:gen_models}.

\subsection{Generative models}
\label{sec:gen_models}

The alternative approaches discussed in the previous section that rely on a ``flow'' transformation require an essential ingredient, i.e., how can we move from one easy or trivial distribution to another which is more difficult to sample from?

Generative models provide a powerful tool to do so: they are designed to generate samples from a target probability distribution, which in the case of lattice field theories is analytically known, generally starting from an analytically tractable prior distribution. The generative model induces on the target space a probability density $q_\theta$ that approximates the target density $p$. The need for exactness typically requires that  $q_\theta$ must be analytically tractable, although other methods of unbiased sampling such as path weights are also feasible, as discussed below.

Crucially, the scaling of the costs changes completely when compared to the one of a standard MCMC, see Eq.~\eqref{eq:mc_cost_total}:
\begin{equation}
 C^{(\rm flow)}_{\rm tot} (N) = C^{(\rm flow)}_{\rm train} + N C^{(\rm flow)}_{\rm s}
\end{equation}
with $C^{(\rm flow)}_{\rm train}$ being the fixed training costs and $C^{(\rm flow)}_{\rm s}$ given in the case of importance sampling by Eq.~\eqref{eq:is_cost_sampling}.

A necessary condition to mitigate critical slowing down is that the sampling cost now scales in the continuum better than a traditional Monte Carlo. However, the costs of training an architecture that features such an improved scaling might be very large compared to the thermalization time or the sampling costs for systems not yet affected by severe autocorrelations, see the cartoon in Figure~\ref{fig:cost-comparison} for a simplified graphical comparison. 
We refer to~\cite{Abbott:2022zsh} for a thorough discussion of the trade-off between training and sampling times and how to perform an honest comparison with MCMC costs. Furthermore, we mention the ability of some generative flow architectures to be generalizable across parameter space, i.e., being trained simultaneously for different values of the model parameters; see~\cite{dibak2022temperature,Gerdes:2022eve,Singha:2022icw,Zhu:2025pmw} for some examples.

\begin{figure}
    \centering
    \includegraphics[width=0.8\linewidth]{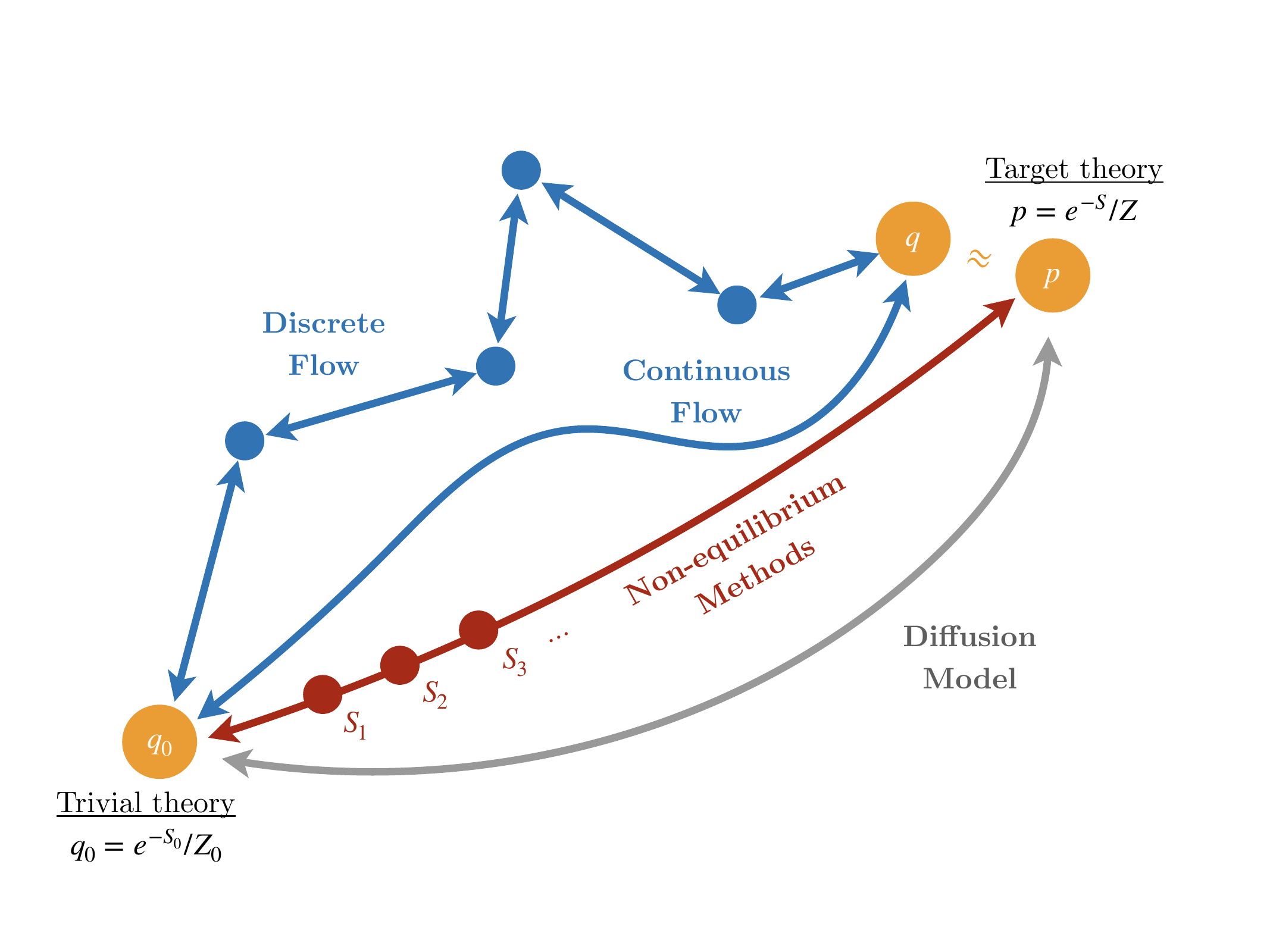}
    \caption{Illustrative comparison of the three transport strategies outlined in the main text.}
    \label{fig:transport_strategies}
\end{figure}

The remainder of this section explores the three broad strategies commonly used to define the flow transformation itself. The distinction between these three approaches is whether and how the flow transformation is constrained to follow specific intermediate distributions between the prior and target distribution:
\begin{enumerate}
    \item General normalizing flow models follow \textbf{no defined path}.
    \item Non-equilibrium approaches follow an \textbf{explicitly defined path}, specified by a changing action.
    \item Diffusion models and related approaches follow an \textbf{implicitly defined path}, specified by a transformation of the field configuration samples.
\end{enumerate}
These approaches are schematically compared in Figure~\ref{fig:transport_strategies}.

\subsubsection{No defined distribution path: normalizing flows}

Normalizing flows (NF)~\cite{tabak2010density,tabak2013family} can be used as a parametrization of the maps discussed in section~\ref{sec:alternative_sampling}, both in the context of flow-based importance sampling and of (approximate) trivializing maps: here we will discuss applications in lattice field theory of both \textit{discrete} NFs (for a variety of implementations see refs.~\cite{Kanwar:2020xzo,Albergo:2019eim,Nicoli:2019gun,Nicoli:2020njz,Boyda:2020hsi,DelDebbio:2021qwf,Albergo:2021bna,Hackett:2021idh,Foreman:2021ljl,Finkenrath:2022ogg,Albergo:2022qfi,Abbott:2022zhs,Singha:2022icw,Abbott:2022zsh,Nicoli:2023qsl,Albandea:2023wgd,Abbott:2023thq,Abbott:2024kfc,Bulgarelli:2024yrz,Schuh:2025gks,Hasenfratz:2026azk,Hasenfratz:2026mae,Singha:2026aac,Singha:2026qyx}), and \textit{continuous} NFs (see refs.~\cite{Gerdes:2022eve,deHaan:2021erb,Bacchio:2022vje,Caselle:2023mvh,Gerdes:2024rjk}).

A discrete normalizing flow is a diffeomorphism $f_\theta$ built as a combination of invertible maps $f_i$, usually denoted ``coupling layers''. In the following, we collectively indicate with $\theta$ the set of parameters defining the transformation $f_\theta$.
In the case of a lattice gauge theory simulation, it takes a field configuration $U_0$ sampled from some prior distribution $q_0$ and gives as output a new configuration $U$:
\begin{equation}
\label{eq:nf_def}
    U = f_\theta (U_0) = (f_n \circ \cdots \circ f_1 )(U_0).
\end{equation}
The training procedure of a normalizing flow requires that the Jacobian of each coupling layer $f_i$ be efficiently computable, so that the determinant of the Jacobian of the whole transformation $f_\theta$ can be calculated as
\begin{equation}
    \det J_{f_\theta} = \det J_{f_1} \cdot \ldots \cdot \det J_{f_n}.
\end{equation}
To achieve invertibility and a triangular Jacobian, the maps $f_i$ usually act separately on two subsets of the field configuration: one is ``active'', i.e. the variables are modified by the map, while the other is kept ``frozen'' and used purely as an input for the map. As a simplified example in lattice gauge theory, we can take
\begin{equation}
f_i:
    \begin{cases}
        U'_{\rm active} = W_{\theta_i} (U_{\rm frozen}) \cdot U_{\rm active} \\
        U'_{\rm frozen} = U_{\rm frozen},
    \end{cases}
\end{equation}
where $W_{\theta_i}(\cdot)$ is a group-valued function.
This transformation is invertible regardless of the form of the function $W_{\theta_i}$, which can be parametrized by arbitrarily deep and complex neural networks. Simple examples of coupling layers are NICE~\cite{dinh2014nice} and RealNVP~\cite{dinh2016density}, which can be readily used in a scalar field theory with an even-odd decomposition. For recent implementations of coupling layers enforcing gauge symmetries, please refer to~\cite{Kanwar:2020xzo,Nagai:2021bhh,Abbott:2023thq} and to~\cite{Albergo:2021bna,Abbott:2022zhs} for gauge theories with fermions. 

Crucially, the probability distribution of the output of a discrete normalizing flow can be explicitly written using the change of variables theorem as
\begin{equation} \label{eq:flow_density}
    q_\theta (U) = q_0 ( f_\theta^{-1}(U) ) \left| \det J_{f_\theta} \right|^{-1},
\end{equation}
so that the reweighting factors $w(U) = \frac{e^{-S(U)}}{q_\theta(U)}$ can be computed when evaluating the path integral of Eq.~\eqref{eq:is_path_integral}.

An alternative architecture to discrete coupling layers is represented by continuous normalizing flows~\cite{Chen:2018wjc,zhang2018monge,zhang2021diffusion}. In this case, the invertible map $f_\theta$ is defined by integrating the ordinary differential equation (ODE)
\begin{equation}
\frac{d}{dt} U(t) = \nabla \varphi_\theta(U(t); t)
\end{equation}
starting from a given configuration $U_0$ and parameterized by the scalar function $\varphi_\theta$. For lattice gauge theories, the gradient $\nabla$ takes values in the Lie algebra, and the resulting field configuration at arbitrary times $t$ may then be written as a path-ordered exponential,
\begin{equation}
U(t) = \mathcal{P}\exp \left\{ \int_0^t dt' \, \nabla \varphi_\theta(U(t'); t') \right\} U_0.
\end{equation}
The output of the map is conventionally defined as the value at $t=1$, i.e., $f_\theta(U_0) = U(t)|_{t=1}$.
Importantly, an analogous ODE can be used to compute the Jacobian determinant as
\begin{equation}
    \log \det J_{f_\theta} = \int_0^1 dt \, \nabla^2 \varphi_\theta(U(t); t),
\end{equation}
giving $q_\theta(U)$ via Eq.~\eqref{eq:flow_density} as for discrete flows.

The choice of the scalar function $\varphi_{\theta}$ parametrizes the flow, and may for example be determined analytically or using optimized neural networks.
The trivializing map of Ref.~\cite{Luscher:2009eq} is a particular parameter-free choice of a continuous flow, for which the flow $\nabla \varphi$ is evaluated in a perturbative expansion to a given order; unfortunately, the convergence of this expansion limits the range of lattice spacings for which this approach is practical.

In order to obtain a flow $f_\theta$ that can be used in a realistic simulation, a training procedure on the parameters $\theta$ must be performed: this is one of the main challenges of flow-based approaches along with the parametrization of $f_{\theta}$ itself. For both discrete and continuous normalizing flows, training is done by minimizing certain metrics that measure the level of similarity between $q_\theta(U)$ and $p(U)$. The most used in the context of NF is the so-called (reverse) Kullback--Leibler (KL) divergence, defined as
\begin{equation}
\label{eq:dkl}
    D_{\rm KL} (q_\theta \| p) = \int \mathcal{D} U \, q_\theta (U) \log \frac{q_\theta (U)}{p(U)} = -\langle \log w (U) \rangle_{U \sim q_\theta} + \log Z \, \geq 0.
\end{equation}
This quantity is minimized when $q_{\theta} = p$, so training to minimize the KL divergence drives the model distribution $q_{\theta}$ towards the desired Boltzmann distribution.
In the last step it is made explicit that this is a case of a \textit{self-training} procedure: indeed, this metric is computed using samples from the model itself and not from the target $p$. Note however that this is not the only possibility, as other loss functions can be evaluated in a self-training approach (and can be varied during training as well). Moreover, another typical choice is using the forward KL divergence and training using samples from an MCMC, or a mix of the two approaches. 

For lattice gauge theories, it has also been found to be helpful to constrain flow models to exactly satisfy certain symmetries, particularly including the high dimensional lattice gauge symmetry. The probability density $q(U)$ of a flow model is guaranteed to be invariant with respect to any symmetry generator $g$ if~\cite{kohler2020equivariant}
\begin{enumerate}
    \item The prior density is invariant, $q_0[g \cdot U] = q_0[U]$, and
    \item The flow is equivariant, $f \circ g = g \circ f$\footnote{The term ``equivariant'' is a property of maps, such as the flow $f$, whereas ``covariant'' is typically reserved for fields with well-defined transformation properties under the action of generators $g$.}.
\end{enumerate}
Several classes of discrete and continuous flows have been developed to be equivariant under gauge and other symmetries~\cite{Kanwar:2020xzo,Boyda:2020hsi,Gerdes:2022eve}.
Outside of normalizing flows, recent works have also developed and applied new architectures for more general gauge-equivariant transformations~\cite{Favoni:2020reg,Nagai:2021bhh,Namekawa:2022liz,Lehner:2023bba,Knüttel:2023w1,Aronsson:2023rli,Lehner:2023prf,Holland:2023lfx}. An open question in this area is how to achieve a good tradeoff between flexibility of the flow $f_\theta$ and the implementation of the constraint to be symmetric, and future improvements may emerge from integrating non-flow methods.

\subsubsection{Explicitly defined distribution path: non-equilibrium approaches}

A similar approach to importance sampling, but purely stochastic in nature, is given by algorithms inspired by Jarzynski's equality~\cite{Jarzynski1997,Jarzynski1997_2} and (equivalently) by Annealed Importance Sampling (AIS)~\cite{Neal2001}. 
In these approaches, one builds between the prior $q_0 \sim e^{-S_0}$ and the target $p \sim e^{-S}$ a sequence of intermediate probability distributions that interpolate between them. The initial configuration $U_0$ is sampled from $q_0$: then it is updated following a pre-determined sequence of intermediate distributions until reaching the target in a finite number of steps. In particular, we have something like
\begin{equation}
\label{eq:noneq_seq}
  U_0 \stackrel{S_{\lambda(1)}}{\longrightarrow} \; U_1 \;
  \stackrel{S_{\lambda(2)}}{\longrightarrow} \; U_2 \;
  \stackrel{S_{\lambda(3)}}{\longrightarrow} \; \dots \;
  \stackrel{S_{\lambda(n_{\rm step})} \equiv S}{\longrightarrow} \;  U
\end{equation}
where each arrow indicates a MCMC update performed using an intermediate distribution, itself defined by the action $S_{\lambda(n)}$, which depends on a (set of) parameter(s) $\lambda(n)$ varied throughout the process.
Since in general we never let the system thermalize at each step, this is a typical example of a non-equilibrium Monte Carlo simulation. Interestingly, such an approach can be used to compute the path integral of Eq.~\eqref{eq:is_path_integral} as
\begin{equation}
\label{eq:noneq_path_integral}
    \langle \mathcal{O} \rangle = \int \left( \prod_{i=0}^{n_{\rm step}} \mathcal{D} U_i \right) \; q_0(U_0) \; \mathcal{P}_{\rm f}[U_0, U_1, \dots U] \; e^{-(\mathcal{W} - \Delta F)} \; \mathcal{O}(U) = \frac{\langle \mathcal{O} e^{-\mathcal{W}} \rangle_{\mathrm f}}{\langle e^{-\mathcal{W}} \rangle_{\mathrm f}},
\end{equation}
where $\mathcal{P}_{\rm f}$ is the transition probability of the sequence of updates of Eq.~\eqref{eq:noneq_seq} and $\mathcal{W}$ the work performed on the system when moving from one intermediate action to the next (with $S_{\lambda(0)} := S_0$),
\begin{equation}
\label{eq:work}
    \mathcal{W} = \sum_{i=0}^{n_{\rm step}-1} S_{\lambda(i+1)} (U_i) - S_{\lambda(i)} (U_i).
\end{equation}
Furthermore, in Eq.~\eqref{eq:noneq_path_integral} we also used Jarzynski's equality, i.e. $e^{-\Delta F} =\langle e^{-\mathcal{W}} \rangle_{\mathrm f}$, with $\Delta F$ being the difference in free energy between $q_0$ and $p$. Notably, $e^{-\mathcal{W}}$ plays the role of the reweighting factor $w$ defined in the previous sections. In the last decade this approach has found applications in lattice gauge theory, both for the computation of partition functions~\cite{Caselle:2016wsw,Caselle:2018kap} and as a tool to mitigate critical slowing down~\cite{Bonanno:2024udh}.

This approach can be augmented with the same tools used to define normalizing flows. Namely, one can redefine the sequence of Eq.~\eqref{eq:noneq_seq} as
\begin{equation}
    U_0 \stackrel{f_1}{\longrightarrow} \; f_1(U_0) \;
  \stackrel{S_{\lambda(1)}}{\longrightarrow} \; U_1 \;
  \stackrel{f_2}{\longrightarrow} \; f_2(U_1) \;
  \stackrel{S_{\lambda(2)}}{\longrightarrow} \; \dots \;
  \stackrel{S_{\lambda(n_{\rm step})} \equiv S}{\longrightarrow} \;  U
\end{equation}
with the $f_i$ being for example the invertible transformations defining the NF in Eq.~\eqref{eq:nf_def}: this defines a stochastic normalizing flow~\cite{wu2020stochastic} which has been extensively implemented in lattice field theory~\cite{Caselle:2022acb,Caselle:2024ent,Bulgarelli:2024brv,Bonanno:2025pdp}; see also Ref.~\cite{Chen:2026bqw} for related architectures. Note that the Continual Repeated Annealed Flow Transport Monte Carlo (CRAFT)~\cite{Matthews:2022sds} approach follows a very similar philosophy. The work is straightforwardly generalized to
\begin{equation}
\label{eq:work_snf}
    \mathcal{W} = \sum_{i=0}^{n_{\rm step}-1} S_{\lambda(i+1)} (f_{i+1}(U_i)) - S_{\lambda(i)} (U_i) - \log \det J_{f_{i+1}} (U_i).
\end{equation}
Unlike NFs, the probability distribution $q(U)$ of the flow is not analytically accessible and the divergence of Eq.~\eqref{eq:dkl} cannot be used for training the parameters. Interestingly, a different divergence can be used to train such architectures, namely
\begin{equation}
\label{eq:dkl_snf}
    D_{\rm KL} (q_0(U_0) \mathcal{P}_f \| p(U) \mathcal{P}_r) = \langle \log \frac{q_0(U_0) \mathcal{P}_f[U_0, U_1, \dots U]}{p(U) \mathcal{P}_r[U_0, U_1, \dots U]} \rangle_{\mathrm{f}} = \langle \mathcal{W} \rangle_\mathrm{f} - \Delta F \geq 0;
\end{equation}
here $\mathcal{P}_r$ indicates the overall transition probability of a sequence of updates similar to Eq.~\eqref{eq:noneq_seq}, but starting from $p$ and reaching $q_0$.
So, the KL divergence of Eq.~\eqref{eq:dkl_snf} measures the similarity between the stochastic flow moving forward (from $q_0 \to p$) and in reverse (from $p \to q_0$): as for NFs this can be computed using only samples from the forward process. The positivity of the KL divergence also implies the last step of Eq.~\eqref{eq:dkl_snf}, which has a clear interpretation in terms of the Second Law of thermodynamics.

Another promising approach can be obtained starting from AIS and non-equilibrium MCMC and taking the limit in continuous time using annealed Langevin dynamics, i.e., solving the Stochastic Differential Equation (SDE)
\begin{equation}
\mathrm{d} \phi_t = - \epsilon_t \nabla S_{\lambda(t)}(\phi_t) \mathrm{d}t + \sqrt{2 \epsilon_t} \mathrm{d}\eta_t, \qquad \phi_0 \sim \rho_0.
\end{equation}
Here $\phi_t$ is the field configuration, $S_{\lambda(t)}$ is the action at time t following a certain protocol for the time-dependent parameter(s) $\lambda(t)$, $\epsilon_t$ is a time-dependent diffusion coefficient, and $\eta_t$ represents a unit-normalized Wiener noise process. Following Eq.~\eqref{eq:work}, the infinitesimal work can then be naturally defined as 
\begin{equation}
    \mathrm{d} \mathcal{W}_t = \partial_t S_{\lambda(t)} (\phi_t) \mathrm{d}t;
\end{equation}
expectation values and the partition function can then be computed as in Eq.~\eqref{eq:noneq_path_integral}.

Exactly as the discretized version, such samplers can suffer from high variance when too far from the equilibrium distribution throughout the transformation. To solve or mitigate this problem, the idea proposed in approaches such as NETS~\cite{Albergo:2024trn} (see also~\cite{vargas2024transport,Kapust:2025ida}) is to introduce an additional transport term. The process now becomes
\begin{align}
    \mathrm{d} \phi_t & = - \epsilon_t \nabla S_{\lambda(t)}(\phi_t) \,\mathrm{d}t + b_t(\phi_t) \,\mathrm{d} t + \sqrt{2 \epsilon_t} \,\mathrm{d}\eta_t, \qquad \phi_0 \sim \rho_0, \\
    \label{eq:nets2}
    \mathrm{d} \mathcal{W}_t & = - \nabla \cdot b_t (\phi_t) \,\mathrm{d}t + \nabla S_{\lambda(t)}(\phi_t) \cdot b_t (\phi_t) \,\mathrm{d}t +  \partial_t S_{\lambda(t)} (\phi_t) \,\mathrm{d}t
\end{align}
where $b_t$ is a velocity field that can be learned. In the perfect transport case, this new field would satisfy: 
\begin{equation}
    \nabla \cdot (b_t \rho_t) = -\partial_t \rho_t
\end{equation}
with $\rho_t = Z_{\lambda(t)}^{-1} e^{-S_{\lambda(t)}}$. It is easy to prove that the weights of Eq.~\eqref{eq:nets2} would simply become
\begin{equation}
    \mathrm{d} \mathcal{W}_t = \partial_t F_{\lambda(t)} \mathrm{d}t
\end{equation}
with $F_{\lambda(t)} = - \log Z_{\lambda(t)}$.

\subsubsection{Implicitly defined distribution path: diffusion models}

Diffusion models provide an alternative framework for generating samples from an empirical probability distribution represented by a dataset. Such models can be essentially described as two distinct processes: in the forward process, noise is incrementally added to the data, while the reverse process (the so-called \textit{denoising}) tries to remove the added noise and recover the (unknown) probability distribution of the starting dataset. The forward process \emph{implicitly} defines a path through the space of distributions $p_t$ that is designed to bridge the target distribution $p_0$ with a simple Gaussian distribution $p_T$. The reverse process is then optimized to approximate this bridge.

The forward process is described by a Langevin process in the time $t = 0 \to T$ 
\begin{equation}
    \mathrm{d} \phi_t = v(\phi,t) \mathrm{d} t + g(t) \mathrm{d} \eta_t
\end{equation}
with drift term $v(\phi,t)$ and time-dependent scalar diffusion coefficient given by the square of $g(t)$.
The reverse process, on the other hand, can be defined via the reversed time variable $\bar{t}=(T-t)$ running from $0 \to T$ and it is given by the SDE
\begin{equation}
\label{eq:dm_reverse}
    \mathrm{d} \phi_{\bar{t}} = -v(\phi,t) \mathrm{d} \bar{t} + g(t)^2 \nabla \log{p}_t \mathrm{d}\bar{t} + g(t) \mathrm{d}\eta_{\bar{t}}
\end{equation}
with the drift term now including $p_t(\phi)$, the probability distribution at time $t$ in the forward
diffusion process. Such SDE can be solved after specifying $v$ and $g$ and, most importantly, determining the gradient of the logarithm of $p_t(\phi)$. If this is possible, then Eq.~\eqref{eq:dm_reverse} can be used to generate new samples directly from the target distribution the initial data were sampled from.

Typically, neither $p_t(\phi)$ nor the required \textit{score} term $\nabla \log p_t(\phi)$ are known. The key for a successful denoising procedure lies then in learning an approximate score function $s_\theta (\phi,t) \approx \nabla \log p_t(\phi)$, which can for example be modeled by deep neural networks. This \textit{score matching} procedure is performed by minimizing a proper loss function~\cite{Song:2021xxx}. Since the learned score is only approximate, the generated samples are not exact by themselves. We can restore exactness by using Metropolis-adjusted annealed Langevin steps (MAALA)~\cite{Zhu:2025pmw}, by tracking path weights along the generative process~\cite{Wang:2026pmv,Chen:2026bqw}, or by using the trained model as an energy-based model that defines the intermediate distributions of annealed sampling methods~\cite{Sexty:2026aym,Kanwar:2026prt}.
Note that, in contrast to other methods presented in this section, diffusion models require data sampled with the target distribution in order to actually perform score matching. For more details on the applications to lattice gauge theories and the connection with Stochastic Quantization see~\cite{Wang:2023exq,Zhu:2025pmw,Aarts:2025lpi,Vega:2025hgz,Alharazin:2026lcb,Aarts:2026zzr,Komijani:2026lan,Tomiya:2026tbc,Tomiya:2026fjk}.

More general families of interpolation between the target distribution and a simple distribution have also been proposed in works on stochastic interpolants~\cite{Albergo:2022iol,Albergo:2023xxx} and the flow matching procedure~\cite{Lipman:2023xxx}. As for diffusions, the implicit paths here are also defined using samples from the target distribution, and the map is designed in a way that provides an efficient training routine for the relevant score function or velocity field.

\subsection{Estimating the partition function}\label{flow-partition}

The use of trivializing maps for sampling has a pleasant side-effect, as it becomes straightforward to evaluate the partition function itself, which is connected to many interesting observables such as the pressure for the computation of the equation of state~\cite{Engels:1990vr}, the free energy of interfaces~\cite{Binder_2011,Caselle:2006dv}, 't Hooft loops~\cite{deForcrand:2000fi} and the entanglement entropy~\cite{Calabrese:2004eu}.

Evaluating $Z$ in an ordinary Monte Carlo is cumbersome and unnatural, precisely because the partition function is not written as the expectation value $\langle \mathcal O \rangle$ of any fixed function $\mathcal O$; typical determinations of (ratios of) partition functions involve calculating the derivative of $\log Z$ with respect to some parameter, and then integrating over the parameter itself. On the other hand, the availability of an (approximate) flow $f$ makes evaluating the partition function straightforward. 

Recall that an approximate trivializing map for an action $S(U)$ is defined by the fact that
\begin{equation}
    Z^{-1} e^{-S(f(U_0))} |\det J| \approx q_0(U_0)\text,
\end{equation}
where $q_0$ is a ``trivial'' distribution whose properties are analytically understood. In particular, the \emph{correctly normalized} probability distribution $q_0(U_0)$ can be evaluated efficiently for any $U_0$.

If the trivializing map were exact, we could extract the partition function $Z$ immediately, simply by evaluating both $q_0$ and $S \circ f$ at the same point $U_0$:
\begin{equation}
    Z = \frac{e^{-S(f(U_0))}}{q_0(U_0) |\det J|^{-1}}\text.
\end{equation}
For trivializing maps which are only approximate~\cite{Nicoli:2019gun,Nicoli:2020njz}, we can still extract the partition function $Z$, but now as an expectation value:
\begin{equation}
    Z = \langle w (U) \rangle_{U \sim q} = \left\langle \frac{e^{-S(U)}}{q_0(f^{-1}(U)) |\det J|^{-1}} \right\rangle_{U \sim q}
    \text.
\end{equation}
Note that in the case where only the unnormalized prior weight $e^{-S_0}$ is known, this formula can still be used to extract the ratio $Z/Z_0$, which is often the interesting quantity to study.

In the case of AIS/SNF approaches, one can use Jarzynski's equality in the same way, with the weight now given by the exponential of work of eqs.~\eqref{eq:work} and \eqref{eq:work_snf}, i.e.
\begin{equation}
    \frac{Z}{Z_0} = \langle e^{-\mathcal{W}} \rangle_\mathrm{f}
\end{equation}
where the partition function $Z_0$ of the distribution $q_0$ has been made explicit for consistent notation.

Flow and flow-related architectures have been applied successfully on a number of phenomenologically interesting studies of partition functions and free energies. An incomplete list includes computations of the equation of state~\cite{Caselle:2016wsw,Caselle:2018kap,Nicoli:2020njz}, numerical studies of effective string theories~\cite{Caselle:2023mvh,Caselle:2024ent}, entanglement entropy and $c$-functions~\cite{Alba:2016bcp,Bulgarelli:2023ofi,Bulgarelli:2024onj} and the critical Casimir effect~\cite{Bulgarelli:2025riv}.

\section{Observable estimators}\label{sec:estimators}

Instead of attempting to gather samples more efficiently, we may attempt to reduce the noise in the observables, which reduces the number of samples that are needed. This section concerns such methods:
\begin{itemize}
    \item \emph{Control variates} learn a total derivative which is correlated with the noise in the observable, and then subtract that noise off explicitly.
    \item Methods which enable the generation of \emph{correlated samples}, e.g., for different action parameters, reduce the noise in observables related to variations of those parameters.
    \item By evaluating expectation values on a \emph{deformed integration contour}, we may modify the signal-to-noise ratio without affecting the physics.
\end{itemize}

One particular ``observable'' has a different form from all others: the partition function (or Helmholtz free energy) itself. Most of the methods discussed in \emph{this} section assume that the observable is an expectation value, evaluated over a probability distribution given by the Boltzmann factor $e^{-S}$. This is not the case for the partition function: nevertheless there are ways to evaluate it, and a particularly neat one that uses machine learning methods---specifically normalizing flows---is discussed in section~\ref{flow-partition}.

Some of the ideas discussed in this section---control variates and deformed contours---operate directly on the level where expectation values are computed. One function of field configurations is replaced by another function which, while related, has more desirable statistical properties. The observables we are most interested in estimating, however, are often not simple expectation values, but are instead to be extracted by performing fits to those expectation values. A common example is the mass, extracted as the asymptotic decay constant of a two-point function:
\begin{equation}
    \langle \mathcal O(\tau) \mathcal O(0)\rangle \sim e^{- M \tau}\text.
\end{equation}
Methods like control variates will seek to improve the estimate of the mass by replacing the product $\mathcal O(\tau) \mathcal O(0)$ with an operator which has exactly the same expectation value, but smaller variance. We might also, however, imagine replacing it with an operator that shares the same quantum numbers, but has greater overlap with the state being studied. Recent work has begun to explore this direction, using gauge-equivariant neural networks to parametrize interpolators for static quark-antiquark systems, optimized for overlap with the ground and excited states~\cite{Bellscheidt:2026rjh,Mayer-Steudte:2026vaf}.

\subsection{Motivation}
Consider computing, in lattice QCD, the nucleon two-point function:
\begin{equation}
C_N(\tau) = \langle N(\tau) \bar N(0)\rangle \sim e^{-M_N \tau}
\text.
\end{equation}
For the purposes of estimating the nucleon mass, we want to evaluate $C_N(\tau)$ for $\tau$ as large as possible (to damp away excited state contributions). How large $\tau$ we can evaluate at depends (given a sufficiently large lattice) on how the statistical errors behave as a function of $\tau$. The \emph{signal}---the value of the correlation function itself---is of course known to decay exponentially with decay constant $M_N$. By inspecting the variance of the propagators that form the two-point function $N(\tau) \bar N(0)$, we may see that it does as well:
\begin{equation}
    \langle |N(\tau) \bar N(0)|^2\rangle
    =
    \langle (q \bar q)^3(\tau) (q \bar q)^3(0)\rangle
    \sim e^{-3 M_\pi \tau}
\end{equation}
The variance of the nucleon two-point function is given by the two-point function of three pions, so the noise (the square root of the variance) decays as $e^{-\frac 3 2 M_\pi \tau}$.
The precise value of this noise, like the precise value of the nucleon two-point function, would require a full Monte Carlo to estimate on any given lattice; however we know its asymptotic behavior by inspection.

\begin{figure}
    \centering
    \includegraphics{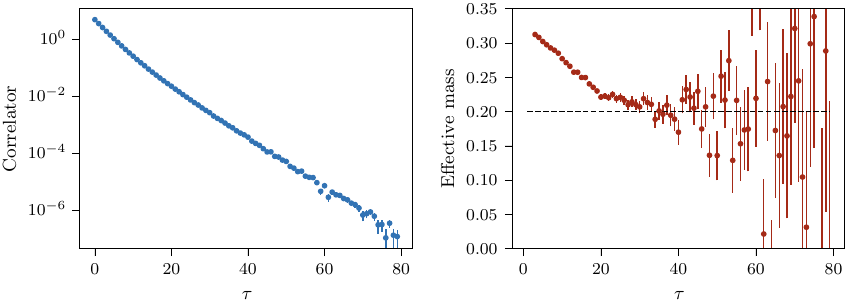}
    \caption{Two views of the signal-to-noise problem, on synthetic data. Left: the correlator on a log-linear plot. The statistical errors grow relative to the signal, although both fall exponentially. Right: an effective mass plot. The smallest mass cannot be effectively determined at early times due to excited-state contamination, and it cannot be determined at late times due to the falling signal-to-noise.}
    \label{fig:stn}
\end{figure}

Although both the signal and the noise decay exponentially, the decay constant for the noise is substantially smaller: $\frac 3 2 M_\pi < M_N$. As a result, the relative statistical errors grow (exponentially) at large $\tau$. This situation is shown in Figure~\ref{fig:stn}.

This \emph{Parisi-Lepage} argument~\cite{Parisi:1983ae,Lepage:1989hd} shows that calculations of the nucleon two-point function suffer in general from a signal-to-noise problem which is exponential in the separation of the source and sink\footnote{One should not confuse this exponential scaling in separation with any sort of statement about asymptotic computational complexity. The task at hand is to determine $M_N$ to within precision $\epsilon$, and this is not exponentially difficult in any parameter of the theory.}, regardless of which operators are used to extract the nucleon state. Analogous arguments indicate that the signal-to-noise problem occurs for almost all correlation functions considered in lattice studies, including higher $n$-point functions.

A closely related problem, the \emph{extensive sign problem}~\cite{Gibbs:1986ut,Splittorff:2006fu,Splittorff:2007ck,Aarts:2008wh,deForcrand:2010ys,Alexandru:2014hga}, occurs when performing Monte Carlo
on a theory with a complex action, such as systems with a non-zero density of relativistic fermions, or real-time dynamics\footnote{In fact sign problems are much more general than this. Almost all quantum Hamiltonians possess sign problems that persist under an arbitrary local change of basis. It is fortunate that many of the systems most interesting in particle and nuclear physics are of the special sort which do not.}. In such systems
the path-integral formula for the expectation value
\begin{equation}
\langle \mathcal O \rangle =
\frac{\int \mathcal D U\, e^{-S(U)} \mathcal O(U)}{\int \mathcal D U\, e^{-S(U)}}
\end{equation}
cannot be interpreted as an expectation value over a probability distribution. To bring Monte Carlo methods to bear on this problem, we most commonly re-express the quantum expectation value as a ratio of classical expectation values taken with respect to the \emph{phase-quenched} system, which is defined by the real part of the action. Concretely, we write
\begin{equation}
    \langle \mathcal{O} \rangle = \frac{\langle \mathcal{O} e^{-i \Im{S}(U)} \rangle_{\Re{S}}}{\langle e^{-i \Im{S}(U)} \rangle_{\Re{S}}},
    \quad 
    \langle X \rangle_{\Re{S}} \equiv \frac{1}{Z_{PQ}} \int \mathcal{D}U \, X(U) \, e^{-\Re{S}(U)}
    \text.
\end{equation}
A little consideration of the properties of the partition functions $Z$ and $Z_{PQ}$ is sufficient to convince yourself that their ratio scales exponentially with the (space-time) volume of the system, and therefore that the denominator of the above expression possesses an exponential signal-to-noise problem, similar to those which afflict correlation functions, but far greater in magnitude.

In most cases, the signal-to-noise problem for correlation functions is \emph{also} due to a sign problem. This follows directly from the Parisi-Lepage argument, which indicates that the average magnitude of measurements of a correlation function is exponentially larger than the mean value to be extracted, which requires strong cancellation of positive and negative sign contributions. This theoretical argument has been complemented by empirical evidence for correlation functions in a variety of contexts~\cite{Beane:2009kya,Beane:2014oea,Wagman:2016bam,Wagman:2017gqi,Davoudi:2020ngi}.

In this text we will primarily focus on these \emph{non-extensive} signal-to-noise/sign problems, arising in systems with a real action.
There are two related reasons for this. First, the extensive sign problem has exponential scaling in the space-time volume of the lattice\footnote{Fermion sign problems scale with the volume measured in physical units (and do not worsen in the continuum limit); real-time sign problems scale with the number of lattice sites (and therefore exhibit exponential scaling in both the continuum and infinite-volume limits).}. For QCD-relevant lattice sizes, we are many orders of magnitude away from making meaningful progress on e.g.~QCD at $T=0$ and large chemical potential. By contrast, signal-to-noise problems have exponential scaling with a length, and are therefore tractable in the near term. Second, the most common computational difficulties encountered in lattice QCD are ordinary signal-to-noise problems. (These are related, of course, because practitioners prefer to work in regimes where we are only a small distance from ``success'' however defined.)

\subsection{Control variates}

We begin with what may be the conceptually simplest approach to a signal-to-noise problem; at least, after getting more statistics. When we wish to evaluate $\langle \mathcal O(U) \rangle$, we will replace $\mathcal O(U)$ with another function $\tilde {\mathcal O}(U)$, which may be proven to have precisely the same expectation value, but has smaller variance. We accomplish this by subtracting from $\mathcal O$ a function $f$, known to have zero expectation. This function is known as a control variate~\cite{Bhattacharya:2023pxx}.

The key observation is that the integral over a compact\footnote{In fact the method does not depend on having a compact configuration space, as long as we're careful that all functions decay sufficiently quickly at large field values---exponential decay is always sufficient.} space, of a total derivative, vanishes. That is, for any function $g$ we have
\begin{equation}
    0 = \int \mathcal D U \,\frac{\partial}{\partial U} \left(g(U) e^{-S(U)}\right)
    = \int \mathcal D U\,
    e^{-S(U)} \left( \frac{\partial g(U)}{\partial U} - \frac{\partial S(U)}{\partial U}g(U)\right)
    \text.
\end{equation}
This is also true for any measure-preserving derivative operator. In what follows we will simply write $\partial$ to stand in for any chosen derivative operator, until we come to the point where the precise operator matters.

Moreover, every (sufficiently regular) function $f$ such that $\langle f \rangle = 0$ can be written in the form $e^{-S} f = \sum_a \partial_a \left(g_a e^{-S}\right)$, summing over a complete set of such derivative operators $\partial_a$. This is vital: it will mean that nothing is lost by restricting our search for control variates to the space of total derivatives. Note that this decomposition is not unique when there is more than one derivative operator $\partial_a$ included in the sum.

With this in mind, we can replace our operator $\mathcal O$ via the equality
\begin{equation} \label{eq:stein-cv}
\langle \mathcal O \rangle = \langle \mathcal O - \partial g + g \partial S\rangle\text,
\end{equation}
for some derivative operator $\partial$.
With this in hand, we are left entirely free to choose $g$ so as to minimize the variance of the improved observable $\mathcal O - \partial g + g \partial S$. 

How do we find a favorable function $g$? An obvious approach is to optimize a neural network~\cite{Bedaque:2023ovz}. We want to minimize the variance:
\begin{equation}
    V[g] = \langle \left(\mathcal O - \partial g + g \partial S\right)^2 \rangle
    - \langle \mathcal O \rangle^2
    \text,
\end{equation}
so we can use $V[g]$ as a loss function.

In practical lattice calculations, the number of configurations available is likely not large relative to the number of parameters in the neural network. If we disregard this and minimize $V[g]$, as defined above, on the $\sim 10^3$ configurations available, we will find that the neural network overfits: it learns a function such that $\partial g - g \partial S$ exactly replicates the fluctuations of $\mathcal O$ on the samples available, but has uncontrolled behavior elsewhere and does not effectively reduce the signal-to-noise problem.

Another approach is to write down a basis of candidate control variates, and optimize a linear combination of these candidates. Given a basis of functions $G_1,\ldots,G_K$, we define the candidate control variates in the same way as above, according to
\begin{equation}
    F_i = \partial G_i - G_i \partial S\text.
\end{equation}
Then the most general linear combination is defined by $K$ coefficients $c_1,\ldots,c_K$:
\begin{equation}
    f = c_i F_i
    \text.
\end{equation}
The chief virtue of this approach is that the minimization of the variance is now a matter of linear algebra. We denote the covariance among the candidates, and the correlations with the observable, as follows:
\begin{equation}
    M_{ij} = \langle F_i F_j \rangle \text{ and } b_i = \langle F_i \mathcal O \rangle
    \text.
\end{equation}
With these definitions, the optimal control variate (in this basis) is given by the coefficients
\begin{equation}
    c = M^{-1} b
    \text.
\end{equation}

Once again, we have a problem when we consider applying this method to a lattice of a size typical in lattice QCD. Optimistically, we might have $10^4$ configurations, on a lattice of size $\sim 50^4$. The number of degrees of freedom in the linear ansatz above is far greater than the number of data points on which we can train.

Regardless of the family of control variates used, it is valuable to ensure that these control variates respect the same symmetries as does the lattice action. To be clear, this is not necessary for correctness---a control variate constructed as described above will have vanishing expectation regardless of its behavior under the various lattice symmetries. However, ensuring that all appropriate symmetries are respected is a cheap way to alleviate overfitting problems.

Control variates have been successfully generalized to both fermionic~\cite{Lawrence:2024xsi} and discrete~\cite{Lawrence:2023cft} systems. The chief obstacle to their application to full-scale lattice QCD is the overfitting problem: it is easy to write down a large family of control variates, but difficult to know which ones are effective without an enormous quantity of training data.
Note also that recent advancements~\cite{Abbott:2026ylv,Abbott:2026vui,Butti:2026gwi} in the use of flow-based approaches as a variance reduction technique naturally lead to a definition of a control variate.

\subsection{Contour deformations}
\begin{figure}
    \centering
    \includegraphics[width=0.4\linewidth]{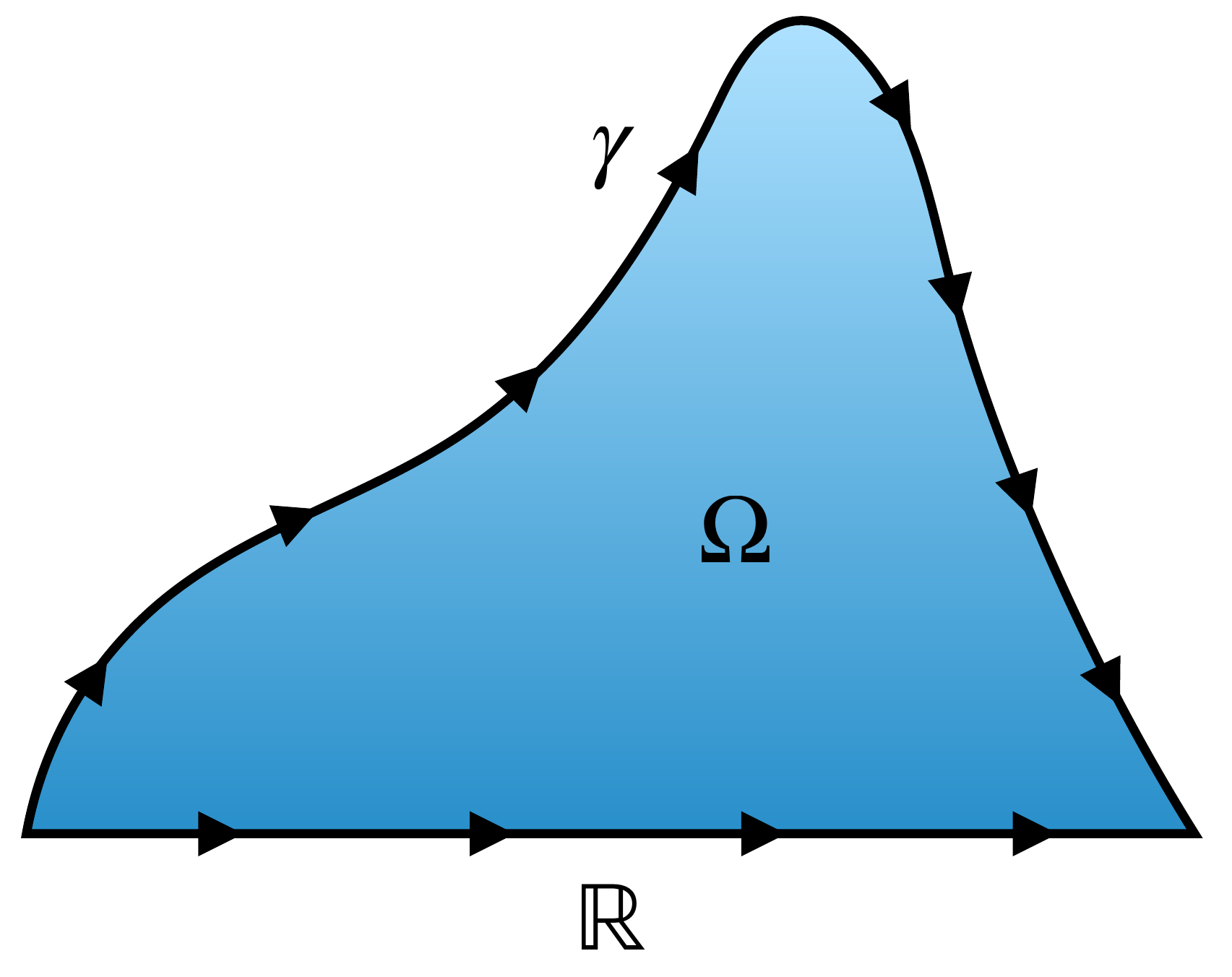}
    \caption{A cartoon of a valid contour deformation. The original domain of integration is the real line, which is then deformed to the curve $\gamma$. Together, the two form the boundary of a region $\Omega$ (symbolically, $\partial \Omega = \mathbb R - \gamma$) on which the function being integrated is holomorphic.}
    \label{fig:cauchy}
\end{figure}

An alternative approach to the signal-to-noise problem starts from complex analyticity\footnote{Or, ``holomorphy.''} of the lattice path integrals. Typical lattice path integrals can be written as (high-dimensional) iterated integrals over real coordinates that can be analytically continued to complex values of the coordinates. When this is the case, Cauchy's integral theorem allows the path integral to be deformed from integrating over the original contour $\mathcal{M}_0$ to a \emph{deformed integration contour} $\mathcal{M}$,
\begin{equation}\label{eq:deformation}
    \langle \mathcal{O} \rangle = \frac{1}{Z} \int_{\mathcal{M}_0} \mathcal{D}U\, \mathcal{O}(U) e^{-S(U)} = \frac{1}{Z} \int_{\mathcal{M}} \mathcal{D}\tilde{U}\, \mathcal{O}(\tilde{U}) e^{-S(\tilde{U})},
\end{equation}
as illustrated in Figure~\ref{fig:cauchy}.

What properties must $\mathcal M$ satisfy in order for equation (\ref{eq:deformation}) to hold? The answer comes from Stokes' theorem, of which equation (\ref{eq:deformation}) is a special case.
The new contour of integration gives an equivalent result whenever the oriented combination of $\mathcal{M}_0 - \mathcal{M}$ bounds a region where the integrand is holomorphic.
In practice, it is sufficient (but not necessary) for $\mathcal M_0$ and $\mathcal M$ to be \emph{homotopic}. That is, it is sufficient for there to exist a connected family of manifolds $\mathcal M_t$, defined on $t \in [0,1]$, such that the integrand is holomorphic all along each manifold.

A practical approach to parameterizing valid integration contours $\mathcal{M}$, is then to define the new contour as the image of a one-to-one map $f_\theta: \mathcal{M}_0 \to \mathcal{M}$. It is sufficient for the map to be connected continuously to the identity map through variations of the parameters $\theta$, as long as this does not traverse any non-analyticities of the path integral. In the context of lattice QCD, the initial integration contour is usually $\mathcal{M}_0 = SU(3)^{4V}$, where $SU(3)$ is viewed as an 8-real-dimensional submanifold of the 8-complex-dimensional space $SL(3;\mathrm C)$. Deformed contours are then submanifolds of $SL(3;\mathrm C)^{4V}$.

Deformed integration contours may be parameterized either analytically, using neural networks, or using hybrid approaches. A tradeoff in the parameterization is that the complex Jacobian determinant $J(U) = \det(\partial \tilde{U}/\partial U)$ is required to integrate on the deformed contour. There is a clear analogy with normalizing flow transformations, and complex normalizing flows have been considered to define deformed contours~\cite{Lawrence:2021izu}. However, unlike the case of real normalizing flows, composition of discrete transformations only gives a valid deformation with a Jacobian of known factorization if each layer is holomorphic, which strongly restricts the available architectures. Hybrid approaches include restricted classes of deformations with constant or diagonal Jacobians~\cite{Alexandru:2018brw,Alexandru:2018fqp,Alexandru:2018ddf,Detmold:2020ncp,Detmold:2021ulb,Rodekamp:2022xpf,Rodekamp:2023byu,Gantgen:2024nut} which can then be optimized using stochastic gradient descent methods. In some cases, determining the parameters of such deformations as the output of neural networks applied to a \emph{constant} input significantly improved the stability of optimization and the final results~\cite{Detmold:2023kjm}.

\subsubsection{Extensive sign problems}
This approach has historically been studied as a possible avenue to circumvent the extensive sign problem~\cite{Cristoforetti:2012su,Aarts:2013fpa,Cristoforetti:2013wha,Mukherjee:2013aga,Aarts:2014nxa,Cristoforetti:2014gsa,Alexandru:2015xva,Alexandru:2015sua,Alexandru:2016gsd,Fujii:2015vha,Tanizaki:2015rda,Alexandru:2017czx,Mori:2017nwj,Tanizaki:2017yow,Alexandru:2018brw,Alexandru:2018fqp,Kashiwa:2018vxr,Fukuma:2019uot,Fukuma:2019wbv,Kashiwa:2019lkv,Mou:2019gyl,Ulybyshev:2019fte,Lawrence:2021izu,Kanwar:2021tkd,Ihssen:2026njd} (see Ref.~\cite{Alexandru:2020wrj} for a review). Here, the challenge is a complex-valued action $S(U)$ which affects both the estimation of the partition function and expectation values of operators. In both cases, Monte Carlo importance sampling according to the absolute value measure $|e^{-S(U)}|$ requires taking expectation values of severely fluctuating complex phase factors $e^{-i \Im{S(U)}}$, for example
\begin{equation}
    \frac{Z}{Z_{PQ}} = \frac{1}{Z_{PQ}} \int \mathcal{D}U \, |e^{-S(U)}| e^{-i \Im S(U)},
\end{equation}
where $Z_{PQ} := \int \mathcal{D}U |e^{-S(U)}|$ is the normalization of the Monte Carlo probability distribution, often referred to as the ``phase quenched'' partition function.

This integral is difficult to estimate by Monte Carlo precisely due to these phase fluctuations. We are computing the expectation value of $e^{-i \Im S}$ over the phase-quenched distribution. Each sample has unit norm; however, the final ratio $Z/Z_{PQ}$ is quite small, and in fact exponentially small in the total space-time volume of the system. We typically refer to $Z / Z_{PQ}$ as the ``average phase'' $\langle \sigma\rangle \sim e^{-m^4 V}$, where $m$ is some mass scale characterizing the severity of this sign problem.

Integrating on a deformed contour has the potential to mitigate this sign problem. The idea is simple: observables are holomorphic functions of the fields, and are therefore not modified (in expectation) by deforming the contour. The average phase, however, is defined by integrands which are not holomorphic, and therefore \emph{can} be changed by the choice of contour. We simply wish to find the contour that reduces the sign problem as much as possible. (Although there are still open questions, a great deal is known today about the nature of such contours; see~\cite{Lawrence:2023sfc} for recent results.)

In practice, integrating on a new contour $\mathcal{M}$ requires a coordinate system to evaluate the integral. It is practical to simultaneously define the contour and coordinate system by applying a diffeomorphic map that sends each original field configuration $U$ to a complexified configuration $\tilde{U}$. Leaving $\tilde{U}$ as an implicit function of $U$, the deformed evaluation of the partition function or an observable can then be written as
\begin{equation} \label{eq:Oave-manif-concrete}
Z = \int \mathcal{D}U \, J(U) \, e^{-S(\tilde{U})},
\qquad 
    \left<\mathcal{O}\right> = \frac{1}{Z} \int \mathcal{D}U \, J(U) \mathcal{O}(\tilde{U}) \, e^{-S(\tilde{U})},
\end{equation}
where $J(U)$ is the (complex) Jacobian of the map from the original integration domain to the deformed contour.

How can we select a contour with a reduced sign problem? We can parameterize the map $\tilde U(U)$ by a neural network, but what should be the loss function? We will use as our loss function the negative of the average sign $\langle \sigma\rangle = Z/Z_{PQ}$, which may be defined on any contour. An obstacle immediately presents itself: this average sign is difficult to evaluate precisely, because of the sign problem. 

A close inspection of the method of gradient descent (and related algorithms) reveals that we do not \emph{need} to evaluate the loss function at any point during the optimization. Instead, we only want to evaluate the gradient of the loss function with respect to parameters. While the evaluation of the loss function $-\langle\sigma\rangle$ suffers from precisely the sign problem we are attempting to solve, derivatives of that loss function do not:
\begin{equation}
    \frac{\delta}{\delta \tilde U(U)}\langle \sigma\rangle[\tilde U(U)]
    = \frac{\delta}{\delta \tilde U(U)} \frac{Z}{Z_{PQ}}
    = - \langle \sigma\rangle \frac{\frac{\delta}{\delta \tilde U(U)} Z_{PQ}[\tilde U]}{Z_{PQ}[\tilde U]}
    \text.
\end{equation}
Here we have used the fact that the partition function $Z$ is contour-independent, and of course we happily disregard an overall factor of $\langle \sigma\rangle$, which does not change the direction of gradient descent. Only the phase-quenched partition function $Z_{PQ}$ changes as the contour changes, and derivatives of the phase-quenched partition function are (essentially by definition) sign problem-free. This approach has been used with a good deal of success on low-dimensional fermionic systems~\cite{Alexandru:2018fqp,Alexandru:2018ddf}. More broadly, neural networks have been used to parameterize contours that reduce the sign problem in a range of systems, including finite-density, Hubbard-model, and real-time problems~\cite{Alexandru:2017czx,Mori:2017nwj,Kashiwa:2018vxr,Kashiwa:2019lkv,Wynen:2020uzx,Lawrence:2021izu,Namekawa:2022liz,Rodekamp:2022xpf,Rodekamp:2023byu}, and normalizing flows have also been used to compute the density of states for complex actions~\cite{Pawlowski:2022rdn}.

\subsubsection{Deformed observables}
Deformations can also be applied in the case that the action is real and the \emph{observable} causes the relevant sign and signal-to-noise problem.
In this context, the aim is to integrate on a contour $\mathcal{M}$ on which the signal-to-noise problem is milder. 
It is also possible in this case to evaluate the deformed path integral by reweighting with respect to the original Monte Carlo samples, because the partition function $Z$ itself has no sign problem. This can be done by defining the new observable
\begin{equation}
Q(U) \equiv J(U) \; \mathcal{O}(\tilde{U}) \; e^{-S(\tilde{U}) + S(U)},
\end{equation}
which has an unchanged expectation value,
\begin{equation} \label{eq:Qave}
\begin{split}
    \left<Q\right> &= \frac{1}{Z} \int \mathcal{D}U \, \left\{ J(U) \mathcal{O}(\tilde{U}) e^{-S(\tilde{U}) + S(U)} \right\} \, e^{-S(U)} \\
    &= \frac{1}{Z} \int \mathcal{D}U \, J(U) \mathcal{O}(\tilde{U}) e^{-S(\tilde{U})} = \left<\mathcal{O}\right>.
\end{split}
\end{equation}


The variance of $\mathcal{Q}$ is in general different from that of $\mathcal{O}$, and the objective is to optimize the manifold parameterization to minimize the variance of the deformed observable. Typically, one is interested in observables that are on average real, in which case the variance $\mathrm{Var}[\Re{Q}] = \left< (\Re{Q})^2 \right> - (\Re\left<Q\right>)^2$ determines the effectiveness of this approach. The second term is unaffected by contour deformations, so a useful choice of loss function to minimize is
\begin{equation} \label{eq:loss-fn}
    \mathcal{L} \equiv \left< (\Re Q)^2 \right>. 
\end{equation}
Both this loss function and its gradient are expressed as an expectation value with respect to the original Monte Carlo distribution, allowing the parameterization to be optimized, and ultimately evaluated, without generating new Monte Carlo samples.
This has been evaluated in a variety of contexts, including complex scalar field theory, Abelian gauge theory, and SU(N) gauge theories in 1+1 and 2+1 dimensions~\cite{Detmold:2020ncp,Detmold:2021ulb,Detmold:2023kjm}.

\subsection{Correlated samples}

For the special case of observables defined in terms of derivatives or differences of primary observables with respect to action parameters, another avenue towards noise reduction lies in the generation of {\it correlated samples}~\cite{Abbott:2024kfc}. Specifically, if two ensembles of gauge‑field configurations are generated at nearby action parameters, $\alpha_1$ and $\alpha_2$, the derivative or finite difference

\begin{equation}
\frac{d\langle O\rangle}{d\alpha} \approx \frac{\langle O\rangle_{\alpha_1} - \langle O\rangle_{\alpha_2}}{\alpha_1 - \alpha_2}\,,
\end{equation}
benefits from potentially substantial cancellation of statistical fluctuations between the two terms when the ensembles are correlated. For highly correlated ensembles, the variance of the estimator may be dramatically reduced compared to both independent sampling at each parameter value and standard finite‑difference reweighting methods. In the limit of a small difference, this has been exactly related to the calculation of control variates of the form of Eq.~\eqref{eq:stein-cv}~\cite{Abbott:2026ylv,Abbott:2026vui,Butti:2026gwi}.
The generation of correlated ensembles can be achieved using the same flow model architectures as described in Sec.~\ref{sec:mc-acceleration} to map \emph{between} theories at nearby points in parameter space.

Physically meaningful applications that involve derivatives with respect to action parameters include chiral and continuum extrapolations, as well as calculations of matrix elements like the chiral condensate, the nucleon sigma term, and other observables using the Feynman--Hellmann theorem. Additional applications arise in evaluating derivatives with respect to the electromagnetic coupling for scale setting or determining isospin-breaking effects in QCD+QED frameworks~\cite{Hermansson-Truedsson:2026zlp}. There is also relevance for theories exhibiting a sign problem, such as computing derivatives with respect to the baryon chemical potential or the QCD $\theta$-angle. In such cases, the distributions connected by the flow transformations are generally more similar than those encountered in sampling-acceleration contexts, and current flow-based techniques are already viable for typical lattice QCD scales.

\section{Observable acceleration}\label{sec:acceleration}

This section is concerned with evaluating a fixed observable as quickly as possible. This is most useful in the case where we have a large number of gauge configurations already gathered, but the observable evaluation is sufficiently slow that we will not be able to evaluate the observable on all gauge configurations, without exceeding the available computational resources.

There are many ways of optimizing the algorithms used to evaluate various observables, and each method inevitably has some free parameters which can be `learned'. To keep this section manageable, we limit the discussion to two particular methods. The first, that of \emph{surrogate observables}, takes a straightforward ML approach, by searching for a neural network (or other model) which approximates the expensive observable. The second, preconditioners, is already in wide use in its conventional (non-ML) form.

\subsection{Surrogate observables}

In the case where we have some observable, a function $\mathcal O(U)$ of the gauge configurations, which is slow to evaluate, we may accelerate things by finding another function $\tilde{\mathcal O}(U)$ which is approximately equal, but can be evaluated more quickly. Obviously, if we have exact equality $\mathcal O(U) = \tilde{\mathcal O}(U)$ for all $U$, we can simply replace the expectation value of the original observable $\mathcal O$ by its faster equivalent. Nothing is lost in such a replacement.

When the equality is not exact, but rather $|\mathcal O - \tilde{\mathcal O}|$ is merely ``typically small'', we can still gain, but more care is required. If we simply replaced $\mathcal O \rightarrow \tilde{\mathcal O}$, we would gain speed but introduce some unknown error. We can handle this error by noticing:
\begin{equation}\label{eq:surrogate-equality}
    \langle \mathcal O \rangle = \langle \tilde{\mathcal O} \rangle - \langle \tilde{\mathcal O} - \mathcal O \rangle
    \text.
\end{equation}
The first term on the right is the fast surrogate for $\langle \mathcal O\rangle$; the second term corrects for the bias introduced.

If we have $N$ configurations, and both right-hand expectation values are evaluated on each of those $N$ configurations, then there is no speed-up from this method: $\mathcal O$ still needs to be evaluated $N$ times. The advantage to (\ref{eq:surrogate-equality}) comes from realizing that there is no need to evaluate both expectation values on all $N$ configurations. If $\tilde{\mathcal O}$ is indeed a good approximation to $\mathcal O$, then the square error $|\mathcal O - \tilde{\mathcal O}|^2$ is small on average, and certainly smaller than the variance of $\mathcal O$ itself\footnote{Note that it is trivial to find a surrogate where the variance in the bias correction is \emph{equal} to the variance in $\mathcal O$: $\tilde{\mathcal O}(U) = 0$ will do the job.}:
\begin{equation}
    \langle \mathcal O^2\rangle
    -\langle \mathcal O\rangle^2
    \gg
    \langle (\tilde{\mathcal O} - \mathcal O)^2\rangle
    -
    \langle \tilde{\mathcal O} - \mathcal O\rangle^2
    \text.
\end{equation}
Therefore, if we are evaluating $\langle \tilde{\mathcal O}\rangle$ on $N$ samples, the number of samples we need to reduce the error in bias-correcting term to be approximately that of the surrogate estimate is
\begin{equation}
\tilde N \sim N \frac{\mathrm{Var}\,({\mathcal O} - \tilde{\mathcal O})}{\mathrm{Var}\,\tilde{\mathcal O}}
    \text.
\end{equation}
In the extreme case where the computational resources required to evaluate $\tilde{\mathcal O}(U)$ are entirely negligible (often a reasonable approximation), the speed-up associated with the use of a surrogate is therefore a factor of $\mathrm{Var}\,(\mathcal O) / \mathrm{Var}\,(\mathcal O - \tilde{\mathcal O})$.

This is reminiscent of control variates, but instead of replacing our observable with one of lower variance, we are replacing it with one which may be more efficiently evaluated. Such a substitution is not designed to reduce our sampling costs, but can reduce computational costs associated with post-sampling measurement.

The reader will hopefully not be surprised at this point by the suggestion that $\tilde {\mathcal O}(U)$ may be learned, rather than guessed by a human. The equations above suggest that the relevant figure of merit for a surrogate observable is
\begin{equation}
    C[\tilde{\mathcal O}(U)] = \mathrm{Var}\,(\tilde{\mathcal O} - \mathcal O)
    \text.
\end{equation}
When the function $\tilde{\mathcal O}(U)$ is smoothly parameterized as in a neural network, this figure of merit is differentiable and therefore usable as a loss function in gradient descent.

This approach has been applied in full-scale lattice QCD calculations only a few times, and with modest success~\cite{Yoon:2018krb,Zhang:2019qiq}. Typically the improvement corresponds to saving $\sim 20\%$ of the total computational effort---substantial, but in practice insufficient to justify the corresponding \emph{human} effort in using these techniques.

\subsection{Preconditioners}

Another promising avenue for accelerating the computation of observables in LQFT involves employing ML techniques to develop improved preconditioners for propagator computations, namely, solutions to the Dirac equation. Preconditioners are transformations applied to linear systems to enhance their numerical properties, facilitating faster convergence of iterative solvers. Formally, given the linear system:
\begin{equation}
D\psi = \eta,
\end{equation}
a preconditioner  is introduced to form a transformed system
\begin{equation}
MD\psi = M\eta,
\end{equation}
which ideally reduces the condition number and thereby accelerates solver convergence.

In lattice field theories, the fields $\psi$ and $\eta$ represent discretized fermionic fields. Conventionally, iterative solvers such as the conjugate gradient (CG) method or BiCGStab~\cite{Frommer:1994vn} are complemented by analytically or heuristically derived preconditioners. Important examples include even-odd (red-black) preconditioning~\cite{DeGrand:1990dk}, domain decomposition methods~\cite{Luscher:2003qa}, incomplete-factorization approaches such as SSOR preconditioning~\cite{Fischer:1996th}, and adaptive multigrid methods (including the closely related low-mode deflation)~\cite{Brannick:2007ue,Luscher:2007se,Babich:2010qb,Frommer:2013fsa,Clark:2016rdz}, which are now the standard choice for light-quark propagator computations. These traditional methods exploit the underlying algebraic and geometric structures of the lattice discretization to substantially enhance solver performance and computational efficiency.

Recently, there has been a growing interest in leveraging ML to create learned preconditioners, which might surpass traditional heuristics through data-driven optimization. While this field remains nascent, early results demonstrate significant potential. Studies have demonstrated that gauge-equivariant neural networks (GENNs) can successfully function as learned multigrid preconditioners~\cite{Lehner:2023prf,Lehner:2023bba,Pfahler:2026fpj}. By explicitly incorporating gauge symmetry---ensuring covariance under transformations $U_\mu(x) \rightarrow \Omega(x) U_\mu(x) \Omega(x+\hat{\mu})^\dagger$---these GENNs respect fundamental symmetries inherent in lattice QCD simulations. Further proof-of-principle research has shown that neural-network-based preconditioners trained on small lattice volumes exhibit robust transferability to larger volumes, suggesting promising scalability and potential computational savings compared to classical methods such as even-odd or incomplete Cholesky preconditioners~\cite{Cali:2022qbd, Sun:2025evk}.

Despite this encouraging progress, several key challenges remain. Ensuring that ML-driven approaches scale efficiently to larger lattices, maintain strict gauge covariance, and generalize effectively across diverse gauge configurations and lattice parameters are ongoing research priorities. Future research directions will likely continue exploring hybrid methodologies, integrating the strengths of classical numerical techniques with innovative ML-driven insights to push forward computational efficiencies in lattice QCD calculations.

\section*{Acknowledgments}

G.K.~was supported by the European Union (ERC, FoundLatt, 101212401). Views and opinions expressed are however those of the authors only and do not necessarily reflect those of the European Union or the European Research Council. Neither the European Union nor the granting authority can be held responsible for them.
S.L.~was supported by a Richard P.~Feynman fellowship from the Los Alamos National Laboratory (LANL) LDRD program; LANL is operated by Triad National Security for the National Nuclear Security Administration of the U.S.~Department of Energy (Contract No.~89233218CNA000001). This manuscript has been approved for unlimited release and assigned the identifier LA-UR-26-29844.
A.N.~acknowledges support by the Simons Foundation grant 994300 (Simons Collaboration on Confinement and QCD Strings), by the European Union -- Next Generation EU, Mission 4 Component 1, CUP D53D23002970006, under the Italian PRIN ``Progetti di Ricerca di Rilevante Interesse Nazionale -- Bando 2022'' prot.~2022ZTPK4E and by the SFT Scientific Initiative of INFN.
PES is supported in part by the U.S. Department of Energy, Office of Science, Office of Nuclear Physics, under grant Contract Number DE-SC0011090 and by Early Career Award DE-SC0021006, and by Simons Foundation grant 994314 (Simons Collaboration on Confinement and QCD Strings), and has benefited from the QGT Topical Collaboration
DE-SC0023646. 
This work is supported by the U.S.\ National Science Foundation under Cooperative Agreement PHY-2019786 (The NSF AI Institute for Artificial Intelligence and Fundamental Interactions, \url{http://iaifi.org/}).

\bibliographystyle{JHEP}
\bibliography{refs.bib}

\end{document}